%% file: publication.tex
\documentclass[a4paper]{article}
\usepackage{hyperref}
\usepackage[english]{babel}
\usepackage[utf8]{inputenc}
\usepackage[colorinlistoftodos]{todonotes}
\usepackage{amssymb,amsfonts,amsmath}
\usepackage[square,sort,comma,numbers]{natbib}
\usepackage{booktabs}
\usepackage{graphicx}
\usepackage{subcaption}
\usepackage[a4paper,margin=2.5cm]{geometry}
\usepackage{amssymb}
\usepackage{arydshln}

\title{Endogeneous Dynamics of Intraday Liquidity}

\usepackage{authblk}
\author[1]{Miko{\l}aj Bi\'nkowski}
\author[2,3]{Charles-Albert Lehalle}

\affil[1]{Department of Mathematics, Imperial College London}
\affil[2]{Capital Fund Management, Paris}
\affil[3]{CFM-Imperial Institute of Quantitative Finance}

\newif\ifdraft
\draftfalse

\def\cmt#1{{#1}} 
\newcommand{\twolines}[2]{
	\begin{tabular}
    	{@{}c@{}}#1 \\ #2
    \end{tabular}
}

\begin{document}
\maketitle 

\begin{abstract}
In this paper we \cmt{investigate} 
the endogenous information contained in four liquidity variables at a five minutes time scale on equity markets around the world: the traded volume, the bid-ask spread, the volatility and the volume at first limits of the orderbook. 
In the spirit of Granger causality, {we measure} the level of information 
\cmt{by the level of accuracy of linear autoregressive models}. This empirical study is carried out on a dataset of more than 300 stocks from four different markets (US, UK, Japan and Hong Kong) from a period of over five years. We discuss the obtained performances of autoregressive (AR) models on stationarized versions of the variables, focusing on explaining the observed differences between stocks.

Since empirical studies are often conducted at this time scale, we believe it is of paramount importance to document endogenous dynamics in a simple framework with no addition of supplemental information. Our study can hence be used as a benchmark to identify exogenous effects.
On another hand, most optimal trading frameworks (like the celebrated Almgren and Chriss one), focus on computing an optimal trading speed at a frequency close to the one we consider. Such frameworks very often take i.i.d. assumptions on liquidity variables; this paper document the auto-correlations emerging from real data, opening the door to new developments in optimal trading.

Thanks to our auto-regressive modeling of these liquidity variables at five minutes time scale, we identify and comment 
stylized facts {in the scope of intraday dynamics}:  a ``\emph{tick size effect}'' {(the smaller the \emph{bid-ask spread in ticks}, the more difficult to predict the bid-ask spread and the easier to predict the volume at first limits using AR models)}, a ``\emph{liquidity effect}'' {(correlation between the market capitalization of a stock and the amplitude of the first coefficient of an AR model for most liquidity variables, independently of the geographical zone)}, and a ``\emph{country-driven effect}'' {(out of sample $R^2$ of AR models on intraday volatility are larger in the US than in Asia, Europe being in between; and a similar geographical ranking for the improvement of using VAR in place of AR models to predict traded value each five minutes)}. 
{Last but not last, we use Granger's $\chi^2$ causality test to identify that valuable information is contained in the past of all four variables to predict each of them: VAR models are more efficient than simple AR models, and the number of informative lags spans from half an hour to more than two hours, depending on the characteristics of the considered stock. We call this property a ``\emph{memory effect}''.}

\end{abstract}



\section{Introduction}
\input{intro.tex}
\section{Methodology and Database}
\label{sec:methodo}
\subsection{Liquidity Driven Variables}\label{variables}

Although liquidity is crucially important for all market participants, the way it might be quantified varies by asset and depends on the investors priorities. As a consequence, a number of liquidity measures has been established within the financial industry (see \cite{citeulike:13704165} for an industry-driven study on liquidity). They attempt to build proxies for transaction costs an average investor is expected to face. As a consequence, they have to reflect the cost of trading a very small number of shares (or contracts) on the one hand, and measure the depth available to large investors on the other hand. For order book driven markets, the former is often measured via the bid-ask spread (the atomic cost of instantaneously buying and selling one share), and the latter is meant to be related to the \emph{volume in the book} (i.e. the average size available to trade at these prices). Since implicit transaction costs\footnote{We will not address \emph{explicit transaction costs} in this paper; they consist of fees, taxes and financing or borrowing costs.} can be mitigated by trading slowly, other metrics refer to the rhythm or speed of the considered market. Traded value and volatility during a reference time interval are typical measures of this kind.  
As a consequence, we have chosen four quantitative measures that seem to be the most important features of liquidity: 
\begin{itemize}
\item \emph{Bid-ask spread} ($\psi$), also known as tightness of the order book, is the difference between the best available ask and bid prices in the limit order book. It describes the cost of completing immediately an order of a number of shares not exceeding the depth on the corresponding first nonempty price limit. 
As we consider 5-minute bins, the bid-ask spread at the start of each bin is taken as the value of the variable.

\item \emph{Traded value/turnover} ($V$), can be defined as the total value of trades carried out within specified period. The big turnovers usually suggest high trading velocity, as well as lower costs of carrying out large orders. Let $\mathcal{B}(d, \tau)$ denote $\tau$-th 5-minute time bin on day $d$.\footnote{i.e. $\mathcal{B}(d, \tau) = [t, t + \Delta\tau), t = d\cdot \Delta d + t_0 + \tau\cdot\Delta\tau$, where $\Delta d =24\mathrm{h}, \Delta\tau = 5\textrm{min}$ and $t_0$ is opening time, dependent on the market.} The traded value is given by
\begin{equation}
 V(d, \tau) = \sum_{t_k \in {\mathcal{B}(d, \tau)}} q(t_k)p(t_k),
\end{equation}


where $(t_k)$ is the series of all consecutive transactions times (date-times), while $q(t)$ and $p(t)$ are respectively traded quantity and traded price recorded at time $t$.\footnote{For convenience, we put $p(t) = p(\max\{t_k: t_k \leq t\}).$}

\item \emph{Volatility} ($\sigma$). From a number of different estimators of this primary feature, we decided to use the \emph{Garman-Klass volatility}, also known as \emph{OHLC volatility}\footnote{Open-High-Low-Close volatlity; in fact, Garman-Klass volatility is the best unbiased analytic scale-invariant estimator based on these four variables, and has $7.4$ higher efficiency relative to na\"ive volatility estimator (in terms of estimator variance). For further information see \cite{GAR80}.}. Note that the microstructure noise in 5-min time-scale has considerable impact on the price variations and therefore the na\"ive estimator would inevitably be biased (see \cite{scales05} for details on the effect of microstructure noise on volatility estimation).

Let $h = \max\{p(t_k): t_k\in \mathcal{B}\}, l = \min\{p(t_k): t_k\in \mathcal{B}\}$ be the high and low traded prices in the bin $\mathcal{B} = \mathcal{B}(d, \tau)$ and let $o = p(\inf\mathcal{B}), c = p(\sup\mathcal{B})$ denote the opening and closing mid-prices for $\mathcal{B}$. The Garman-Klass volatility estimator for the bin $\mathcal{B}$ is given by
 \begin{equation}
  \hat{\sigma}^2(d, \tau) = \tfrac{1}{2}\log^2\left(\tfrac{h}{l}\right) - (2\log(2) - 1)\log^2\left(\tfrac{c}{o}\right).
 \end{equation}

\item \emph{Book size} ($B$) is the amount of shares available at the best bid and best ask levels. The depth of the first-level queue shows roughly how long it is needed to wait for a new limit order to be executed and hence enables a trader to asses the trade-off between the price and immediacy of transaction. It also shows how big market orders can be carried out at the best available price, without additional costs of consuming liquidity on higher/lower price levels.

\end{itemize}

Other quantities of interest include the \emph{Free Float value}, i.e. the share of stocks owned by public investors, and the \emph{average bid-ask spread in ticks} $\psi^*$, i.e. the ratio of bid-ask spread and tick size. The latter is important as it lets us distinguish between \emph{small-tick} stocks ($\psi^* \gtrsim 2 $) and \emph{large-tick} stocks ($\psi^* \lesssim 1.3$). The difference between the distributions of asset returns of small and large-tick stocks is widely known (e.g. \cite{LilloCurato2015}), therefore we shall be interested in the potential differences and dependencies between the aforementioned variables in the context of spread/tick relation.

\subsection{Dataset Description}

The dataset used in this work comes from the recording of market data made by Capital Fund Management (CFM) for research purposes. The original data consists of tick-by-tick records from real-time data feeds connected to each trading venue, stored in binary format and then, when needed, consolidated according to a timestamp provided by each trading venue, recording the time at which the data has been emitted by the venue. Using these raw data, we rearranged them in five minute bins to compute our four liquidity variables.

Only data coming from the continuous auction phases have been used. It means we excluded the large trades of call auctions and the published inconsistent bids and asks during the pre-auction phases (at the open and at the close for all of them, and around the lunch break in our two Asian markets (before and after this break on the Tokyo Stock Exchange and at the start of this break on the Hong Kong exchange%
). We also do not use any data from the lunch breaks present at Asian markets. As a consequence, we have roughly 80 bins in US, 100 bins in the UK, and 60 bins in Asian markets, per day (see \cite{citeulike:12047995} for details on each market).
Half-days of trading (like Christmas) are excluded from the database, but late opening days are included. When there is a trading halt, the reopening call auctions are not taken into account, and bins with no trades (waiting for the re-opening) are excluded, but we can have bins that have been opened to trading only for few seconds or minutes\footnote{if a circuit-breaker is activated and the market has be reopened and is ready to trade few seconds before the end of a bin.}. We have no way to identify explicitly circuit breakers, as a consequence we cannot provide any statistics on their activation.

For each stock, each bin is labelled by the day and the time of the day; it allows to make intraday medians of Figure \ref{fig:AZN:curves}, and to study the intraday seasonality.
Our dataset covers period of January 2011 to March 2016;
Note this is posterior to the implementation of Reg NMS (and the subsequent decimalization, see \cite{hendershott2011does}) in the US and MIFID (and the associated fragmentation, see \cite{citeulike:12047995}) in Europe.
Table \ref{t:datainfo} provides descriptive statistics with a breakdown by country; the main remark is that despite a large range of bid-ask spread in basis points, the bid-ask spread in ticks is less diverse from one zone to another, with Hong-Kong stocks being mainly ``large tick stocks''.
Moreover, we have more US stocks (202) that UK ones (98), and fewer Hong-Kong and Japan stocks (51). To account for microstructure and regulation diversity, we will most often provide breakdowns by country.

For Section \ref{sec:likeffect}, we needed capitalisation of listed companies. That for we used average capitalisation over the period, and the dataset we use are the one provided by the exchanges, that have been recorded by Capital Fund Management.


\begin{table}[!h]
  \centering
  \begin{tabular}{lrrrr}
    \toprule
    location &         Hong Kong &        Japan &        UK &     US \\
    \midrule
    \vspace{3pt}
    total number of stocks & 11 & 40 & 98 & 202 \\
    avg. number of days &  966 &  934 &  1101 &  1241 \\
    avg. total number of 5-minute bins (datapoints) &  63,730 &  55,890 &  110,400 &  96,780 \\
    max. number of bins per day &  66 &  60 &  102 &  78 \\
    \vspace{3pt}
    avg. number of bins per day         &  65.94 &  59.81 &  100.3 &  77.96 \\
    avg. bid-ask spread (bp)   			&  17.48 &  19.30 &  6.76 &       3.55 \\
    avg. bid-ask spread (ticks)         &  1.10 &  1.44 &  1.54 &       1.71 \\
    avg. GK volatility (bp$^2$/bin)  	&  9.45 &  12.44 &  8.63 &       11.46 \\
    \hspace{135pt}\small{(large-tick stocks)} &  9.45 &  12.47 &  7.57 &    12.39 \\
    \hspace{135pt}\small{(small-tick stocks)} & - &  12.17 &  9.27 &      10.93 \\
    avg. book size (\$ million)         &  5.16 &  89.32 &  5.95 &  0.09 \\
    \hspace{135pt}\small{(large-tick stocks)} &  5.16 &  167.30 &  10.51 &  0.12 \\
    \hspace{135pt}\small{(small-tick stocks)} & - &  28.42 &  2.02 &  .054 \\
    avg. book size (in avg. trade size) &  52.3 &  89.7 &  18.7 &      21.3 \\
    \hspace{135pt}\small{(large-tick stocks)} &  52.3 &  192.0 &  28.3 &      31.9 \\
    \hspace{135pt}\small{(small-tick stocks)} & - &  17.3 &  13.2 &      4.7 \\
    avg. turnover (\$ million) 			&  9.73 &  186.62 &  38.99 &       1.20 \\ 
    \hspace{135pt}\small{(large-tick stocks)} & 9.73 &  193.10 &  44.05 &       0.89 \\
    \hspace{135pt}\small{(small-tick stocks)} & - &  93.61 &  23.01 &      2.48 \\
    \bottomrule
  \end{tabular}
  \caption{Dataset statistics per location. Values in all but first row are averages across all stocks per each country. Note that all of the available stocks from Hong Kong were large-tick stocks.}
  \label{t:datainfo}
\end{table}

\subsection{Preprocessing}

\paragraph{Data cleaning.}
In the preprocessing stage, after estimation of all four variables over 5-minute bins, data normalization was carried out. 
Firstly, for each of the stocks, the trading days for which less than 80\% of the total number of bins was available\footnote{Bins might be unavailable due to trading halt or technical reasons.} were completely removed.
Secondly, the zero values for volatility (i.e. when there is no trade during 5 minutes, the volatility is by definition zero) have been substituted with the minimum value higher than threshold $\epsilon=10^{-6}$ observed within the same or the previous day (separately for each stock). This ensured that all of the variables were positive-valued.

\input{stationarize.tex}

%

\subsection{Metrics and Methodology} \label{s:methodology}
We aim to assess the endogenous information in the scope of the four  aforementioned market features. To understand how the information is propagated within market, we will use univariate (AR) and multivariate (VAR) linear autoregressive models. Although these classes of models are basic and their capacity to reflect the relationships between the variables may be limited, they are well studied and commonly used in financial modeling. Ultimately, they let us establish the benchmark lower estimate for the endogenous information, measured as proposed in the above section. 

\paragraph{{Metrics of performance and model selection}.}
Since the preprocessed data is in a time series form, the estimation of an autoregressive model requires selection of lag (the model's order), which is often carried out using \emph{Akaike} or \emph{Bayesian Information Criteria}\footnote{See \cite{aic} and \cite{bic}.}. However, we aim to measure what can be predicted using the past data rather than to find the model that is statistically most reasonable. Therefore, we treat the number of model parameters as a secondary factor, and compare the performance of all of the models of lags up to 40. For each stock and variable, the best model is being selected in accordance with the out-of-sample $R^2$ statistic for the variable's prediction. Note that the VAR model normally would achieve different $R^2$ for each of the variables it comprises.

\paragraph{{Estimation procedure}.}
In the univariate approach we used standard \emph{maximum-likelihood estimator} of the AR model,
separately for each variable. In the multivariate case, VAR models were estimated with all possible subsets of variables. For each prediction task\footnote{Single prediction task was specified by a stock and a variable. Note that for VAR models, each combination of variables count as one prediction task.} the best (subset, lag) pair\footnote{ Following machine learning naming, each such \emph{pair} will be refereed to as the \emph{hyperparameters} of our model.} was selected. Figure \ref{f:frequent_models} presents how frequently each subset of input variables achieved the best results, while Figure \ref{f:frequent_lags} shows the most frequently chosen lags.

To reduce variance of the estimator and standardize the estimation procedure for time-series of different lengths, models were trained and validated separately on 20 partially overlapping batches. Each of the batches contained 150 consecutive training days (i.e. between 9,000 and 15,000 consecutive bins) and another 150 following days that were used for validation. The starting points of the batches were equally distributed, so the first batch contained beginning of the available period, while the last one - the end of the period.

\section{Stylized facts and benchmark results}
\label{sec:illustr}
The linear autoregressive models form standard and well studied class of predictive models that are able to capture linear dependence between variables. 
Their simple yet powerful form makes them usual first candidate for the benchmark solution in modeling time series data. 

\begin{figure}[h!]
   \centering
   \includegraphics[width = 1.\textwidth]{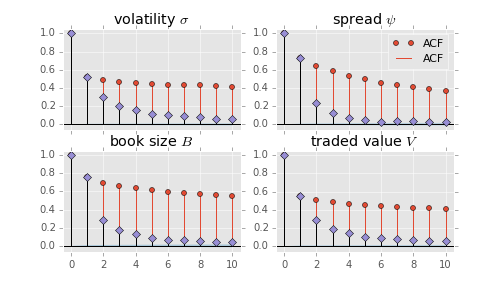}
   \caption[default caption]{Autocorrelation (blue diamonds) and partial autocorrelation (red circles) functions of each of the variables for Welltower Inc. }
   \label{acf}
\end{figure}

After preprocessing that eliminated seasonality and transformed positive-valued variables into real-valued ones, the data has appropriate form for autoregressive modeling. For that approach to make sense, the time series need to exhibit nonzero autocorrelation structure. Although these vary considerably, the positive autocorrelation can be observed for all stocks and variables; an example is presented in Figure \ref{acf}. 

\subsection{Univariate autoregressive modeling}

\begin{table}
\centering
\begin{tabular}{lcccc}
\toprule
{\bf $R^2$ of AR models} &                    US &                       UK &                       Japan &                        Hong Kong \\
Variable & & & & \\
\midrule
volatility          &  \twolines{0.448}{(0.045)} &  \twolines{0.208}{(0.099)} &  \twolines{0.143}{(0.047)} &  \twolines{0.102}{(0.026)} \\\hdashline
avg. bid-ask spread    &  \twolines{0.523}{(0.129)} &  \twolines{0.380}{(0.159)} &  \twolines{0.195}{(0.098)} &  \twolines{0.096}{(0.054)} \\ \hdashline
book size &  \twolines{0.640}{(0.173)} &  \twolines{0.535}{(0.104)} &  \twolines{0.440}{(0.154)} &  \twolines{0.659}{(0.075)} \\ \hdashline
traded value           &  \twolines{0.408}{(0.090)} &  \twolines{0.205}{(0.084)} &  \twolines{0.130}{(0.076)} &  \twolines{0.116}{(0.056)} \\
\bottomrule
\end{tabular}

\caption{Results of univariate prediction across different markets and variables. The presented values are the average $R^2$'s and standard deviations of $R^2$ (in parentheses) of the best models on out-of-sample validation sets across stocks in each market. }
\label{t:results}
\end{table}

The resulting $R^2$ statistics varied across the variables and stocks;
Table \ref{t:results} presents prediction results averaged for different markets. In general, the endogenous information increases with the size of the market. Each of the considered liquidity variables can usually be predicted most accurately for US stocks, followed by those from UK, and ending on Asian ones. The exception can be observed for the book size of the Hong Kong stocks, which can be predicted with $R^2$ of $65.9\%$, which is very close to US stocks ($R^2 \approx 64\%$).

Volatility appears to be the variable for which the achievable predictive power (i.e. $R^2$) is the most stable across the stocks on each market. The markets, however, vary between each other and the average predictability is on different levels. It is worth to notice that the UK stocks have more variable endogenous information and their $R^2$ spans between 0 and $40\%$. More details will be presented in the section \ref{sec:zone_effect}.

Another factor that distinguishes volatility from the other variables is the market memory. As our approach involved the grid search of the best performing model across different hyperparameter settings,\footnote{A single setting was determined by lag number and set of explanatory variables, see Section \ref{s:methodology}.} the lag of the model indicates how far in the past a significant information can be found. The US and UK stocks models required considerably less lagged values to achieve the best performance for volatility than for the other variables\footnote{Moreover, the multivariate models usually did not improve the volatility performance, on the contrary to the other features. See sections \ref{sec:memory_effect} and \ref{s:multivariate} for more details.}.  

The limit order book dynamics is dependent on the tick size of the particular stock\footnote{Let us recall that we consider a stock as a \emph{small-tick stock} ( \emph{large-tick stock}, respectively) if its avg. spread in ticks $\psi^* \gtrsim 2 $ ($\psi^* \lesssim 1.3$, respectively.).}. This has a direct effect on spread prediction itself, as well as on a book size. The explained variance of the book size is approximately inverse-proportional to the spread in ticks, and hence is high for the large-tick stocks and low for small-tick stocks.

\subsection{Correlation analysis.}

\begin{figure}[h!]
   \centering
   \includegraphics[width = .6\textwidth]{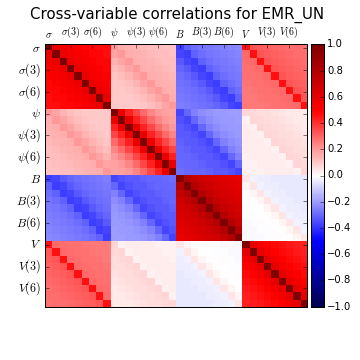}
   \caption[default caption]{Pearson correlations between the variables and their (first nine) lagged values.}
   \label{cross-correl}
\end{figure}

Analysis of the correlations between the features observed at the same time or with a \emph{delay} naturally extends the univariate approach. The positive correlation of volatility and traded volume is a known stylized fact (See \cite{ContEmpiricalProperties})
In this section we consider the dependencies between the market features. Figure \ref{cross-correl} presents the correlations between variables and their delayed values for a sample stock. In general, the cross-variable correlation structure varies considerably by stock. Nevertheless, the strong positive correlation is often observed between volatility and traded volume, while the book size is negatively correlated with the other features.

The correlation structure between time series is closely related with Granger causality. Let us recall that the time series $X$ \emph{G-causes} time series $Y$ if the vector-autoregressive model 
\begin{equation}
	X_t = \sum_{i=1}^p \left(A_{t-i} X_{t-i} + B_{t-i} Y_{t-i}\right) + \epsilon_t
\end{equation}
fits the data \emph{more closely} than the univariate AR model
\begin{equation}
	X_t = \sum_{i=1}^p A'_{t-i} X_{t-i} + \epsilon'_t,
\end{equation}
where $p$ is the lag, $A_i, B_i, A'_i$ are the estimated parameters and $\epsilon_t, \epsilon'_t$ are the error terms. The \emph{goodness of fit} is measured in that case by the variance of the error terms, which is linked directly with the $R^2$ statistic\footnote{The statistic is given by $R^2 = 1 - VAR(\epsilon) \slash VAR(X)$. Since the variance of data does not depend on the model, there is one-to-one correspondence between these performance measures.}. 

Therefore, the comparison of multivariate and univariate models performance will enable us to better understand the relationships between the variables in the context of causality. Note, that the only difference between our approach and classical approach to Granger causality is that we choose the best model parameters in accordance with the out-of-sample $R^2$ instead of the in-sample. 

\subsection{VAR (Vector Auto-Regressive) model}
\label{s:multivariate}
%
%
%

Despite the fact that the considered features measure  incomparable quantities represented in different units, these variables are correlated and as will be shown later, modeling them together using simple linear model can improve prediction of their future values.

In general, we can observe that the VAR models achieved slightly higher explained variance than univariate models in US and UK, yet the improvement was rarely significant and sometimes not present at all. For instance, the AR models of volatility usually outperformed VAR models. This means that volatility time series often possesses all the endogenous information needed to predict its future, and the rest of variables do not cause volatility in the sense of Granger. Interestingly, this is not the case for the other features.  

\begin{figure}[h]
   \centering
   \includegraphics[width =\textwidth]{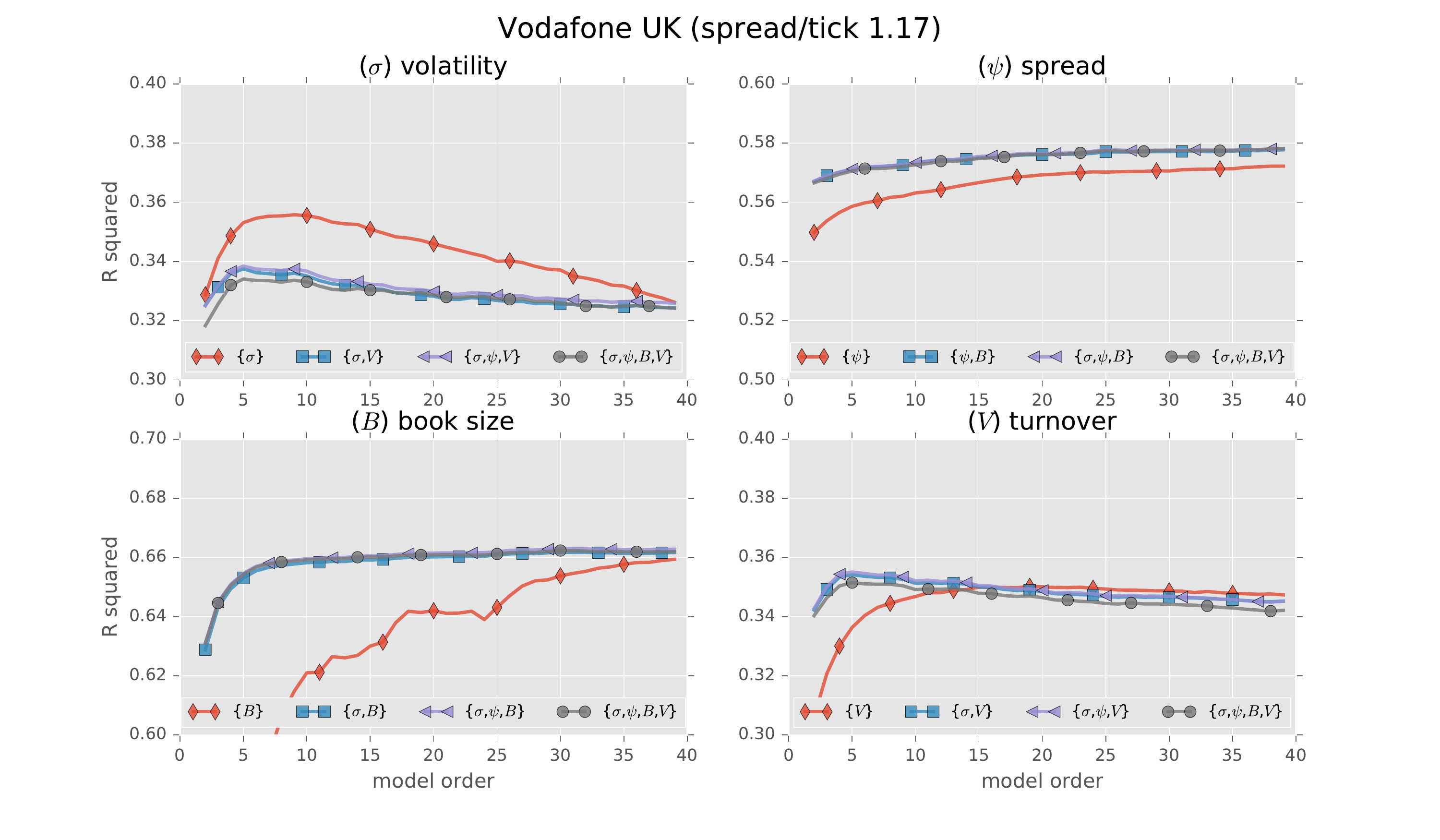}
   \caption[default caption]{Sample \emph{performance curves} for large-tick Vodafone UK stock. Each of the plots presents the $R^2$ statistic for one of the variables, achieved by different AR/VAR models with increasing lags. Each line corresponds to a certain set of explanatory variables. The higher-lag models overfit volatility and turnover, but perform well for spread and book size. Note the small differences between multivariate models. }
   \label{f:learning_curve}
\end{figure}

On the other hand, in many cases the performance close to the univariate benchmark can be achieved by multivariate models of considerably smaller lag. This effect is especially vital for book size and turnover in US and UK, and is presented in the Figure \ref{f:learning_curve}.

There were, however, situations where VAR model yielded better performance than univariate. For most of the stocks in Hong Kong and a group of large-tick stocks in Japan more accurate predictions of turnover and book size were obtained. Lag improvement was also more common on the Asian markets.

\section{Identified effects: New Stylized Facts on Liquidity}
\label{sec:results}

\input{liquidity.tex}

\subsection{Tick size effect}

In this section we discuss \emph{endogenous information} in the context of \emph{avg. bid-ask spread in tick size}. We define the former as the explained variance -- the $R^2$ statistic -- achievable by an autoregressive model in prediction of the future values given the variable immediate past. The latter variable is often used to segment stocks according to the role the bid-ask spread is playing in the microstructure of their limit order books. For the large-tick stocks the value of average bid-ask spread to tick size is small (lower than $1.3$) which means that often there is no empty price limit between the best bid and best ask prices, as well as among several higher levels. This increases the importance of book size as a measure of liquidity and market significance of any price movement.

For small-tick stocks, on the contrary, the spread often stretches over several price levels, making the book size less significant, as liquidity providers submit orders on multiple price levels on each side. The limit orders might also be submitted in between best-bid and best-ask prices, forming a new best-level.

The endogenous information of individual stocks is dependent on the \emph{average bid-ask spread in ticks} and particular variable. 

For book size, the informational content of the past is highest for the large-tick stocks, with $R^2$ between $60\%$ and $90\%$. The statistic is negatively correlated with the spread in ticks but stabilizes at $20$ to $40\%$ for very small-tick stocks. 

\begin{figure}[h]
   \centering
   \includegraphics[width =\textwidth]{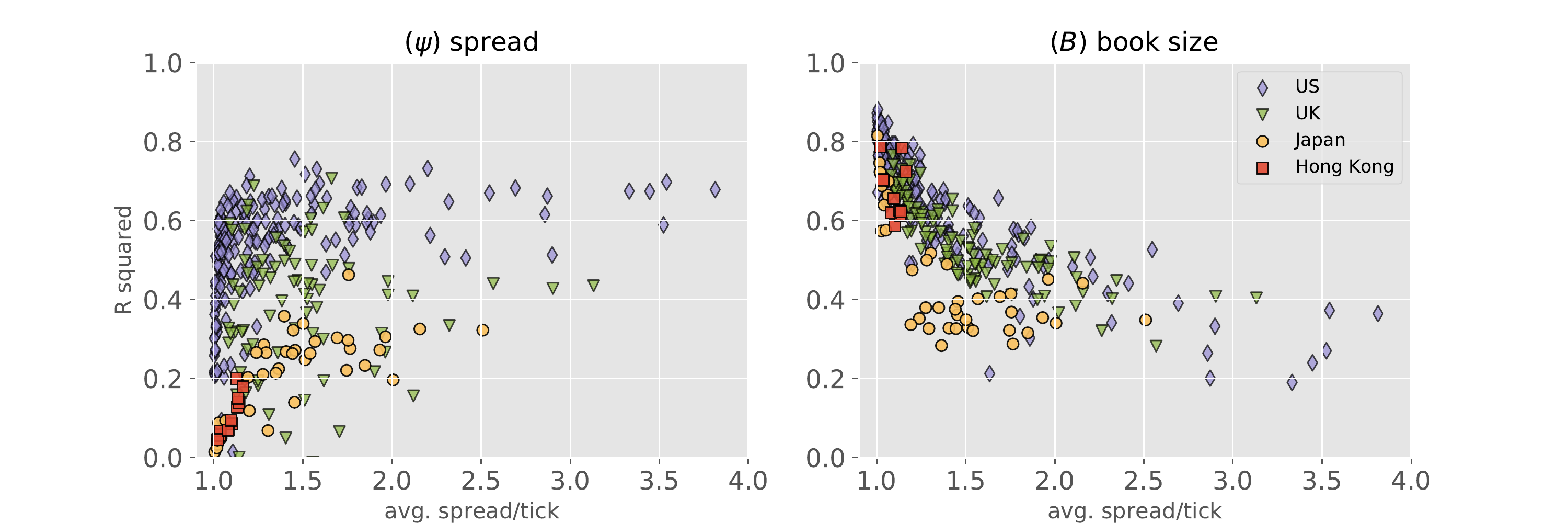}
   \caption[default caption]{$R^2$ in out-of-sample prediction of book size and spread across different stocks from four markets. Each point represents a single stock while it's position indicates the values of achieved $R^2$ and average bid-ask spread in ticks.}
   \label{f:spread_tick}
\end{figure}

All other variables show quite similar patterns. At each market, medium- and small-tick stocks had similar levels of predictability, and higher than the ones of large-tick stocks, for which $R^2$ stretched from $0$ up to $70\%$.

\begin{figure}[!h]
   \centering
   \includegraphics[width =.75\textwidth]{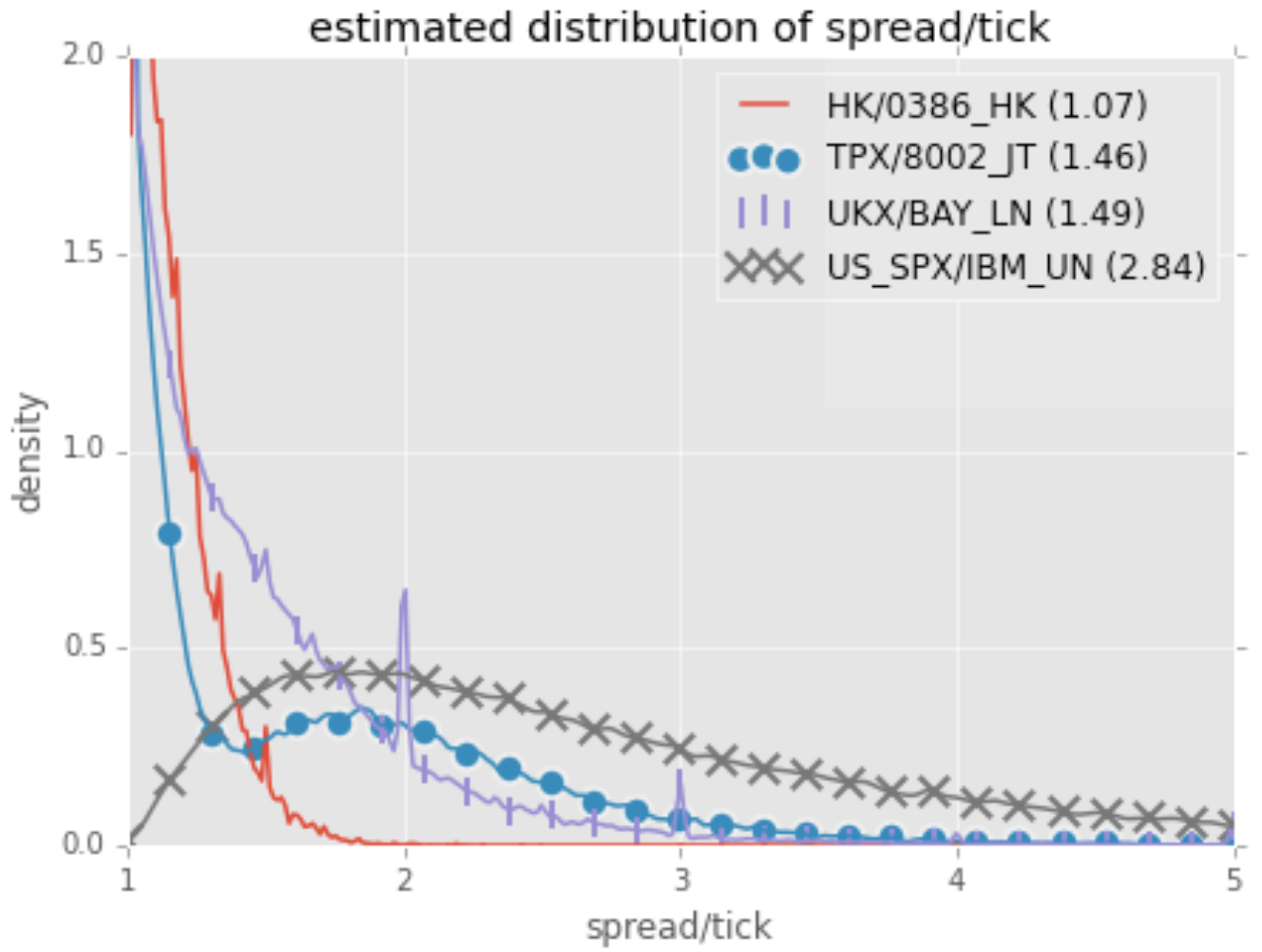}
   \caption[default caption]{Sample distributions of the spread/tick for four different stocks. The stock's spread distributions have discrete probability mass, mostly at $1$, that is decreasing with the increase of avg. spread/tick. For the presented stocks it is equal to .44, .34, .14, .01, respectively. The number next to the stock name shows the average spread/tick size. }
   \label{f:spreadtick_distr}
\end{figure}

These outcomes can easily be observed in Figure \ref{f:spread_tick}. Note that for the average bid-ask spread of large-tick stocks, its poor predicting power might seem counter-intuitive, as the spread for such stocks for most of the time is equal to one tick. This, however indicates that the distribution of spread, averaged over 5-minute bins, is less continuous and has significant mass at $1$ (see Figure \ref{f:spreadtick_distr}), which limits the modeling capabilities of linear regression models.

Moreover, as expected: the large-tick stocks have less volatile bid-ask spread. This has crucial impact on $R^2$, as this statistic is relative to the data deviations; namely, the less variance to predict, the more accurate the predictions must be to retain the given level of $R^2$. Hence, the large-ticks stocks may have their spread predicted more accurately in absolute terms, despite achieving considerably low $R^2$.

To sum up, the large-tick stocks have diverse informational content of the immediate past, generally smaller for bid-ask spread and larger for book size, as compared with the medium- and small-tick stocks of the same market.

\subsection{Country-driven effect}
\label{sec:zone_effect}

\begin{figure}[h]
\begin{minipage}{.48\textwidth}
  \flushleft
  \includegraphics[width=1.08\linewidth]{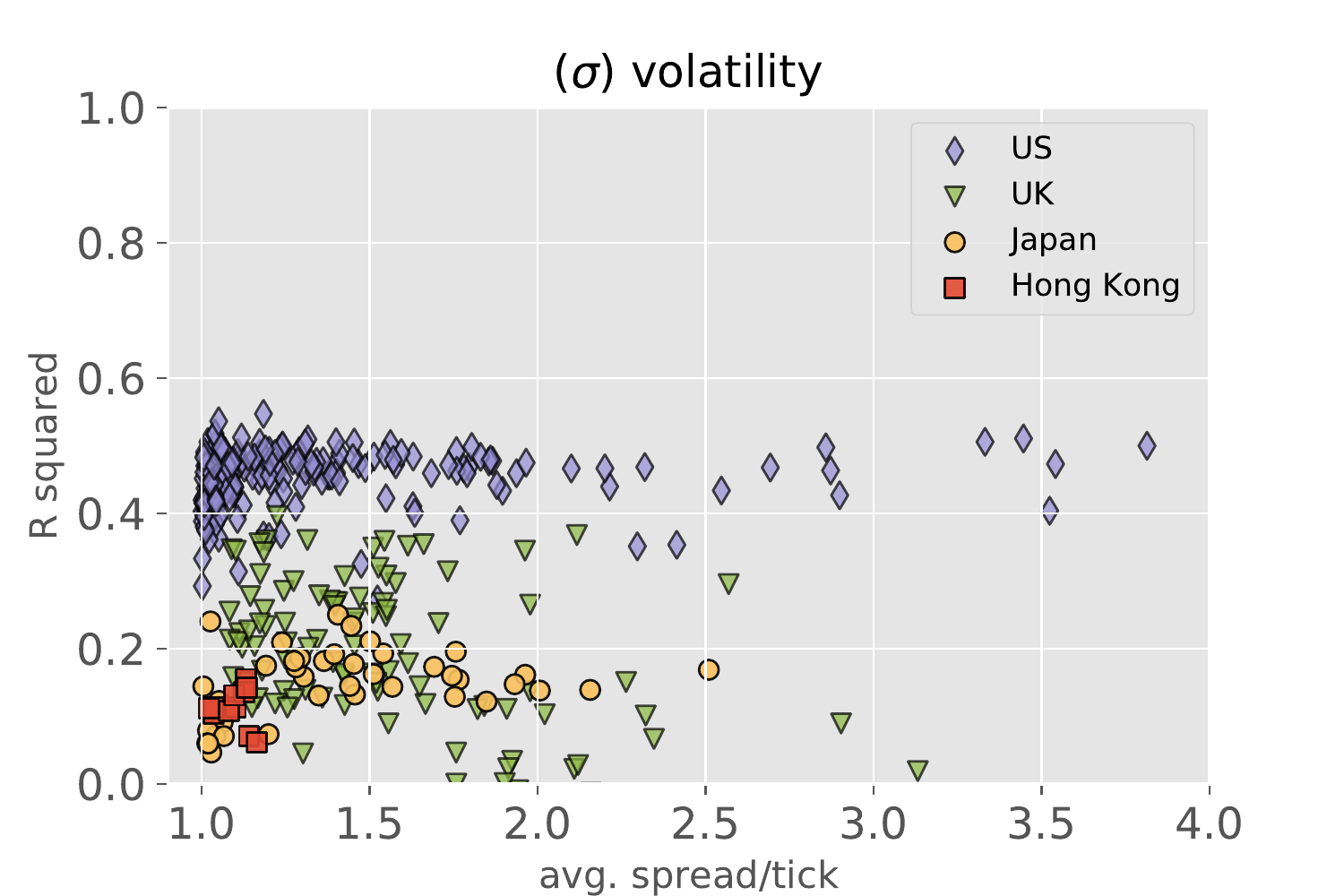}
   \caption[default caption]{$R^2$ in out-of-sample prediction of volatility across different stocks from four markets. Each point represents a single stock's relationship between average bid-ask spread in ticks and informational content of the immediate past. }
   \label{f:spread_tick2}
\end{minipage}
\hfill
\begin{minipage}{.48\textwidth}
\flushright
  \includegraphics[width =1.08\linewidth]{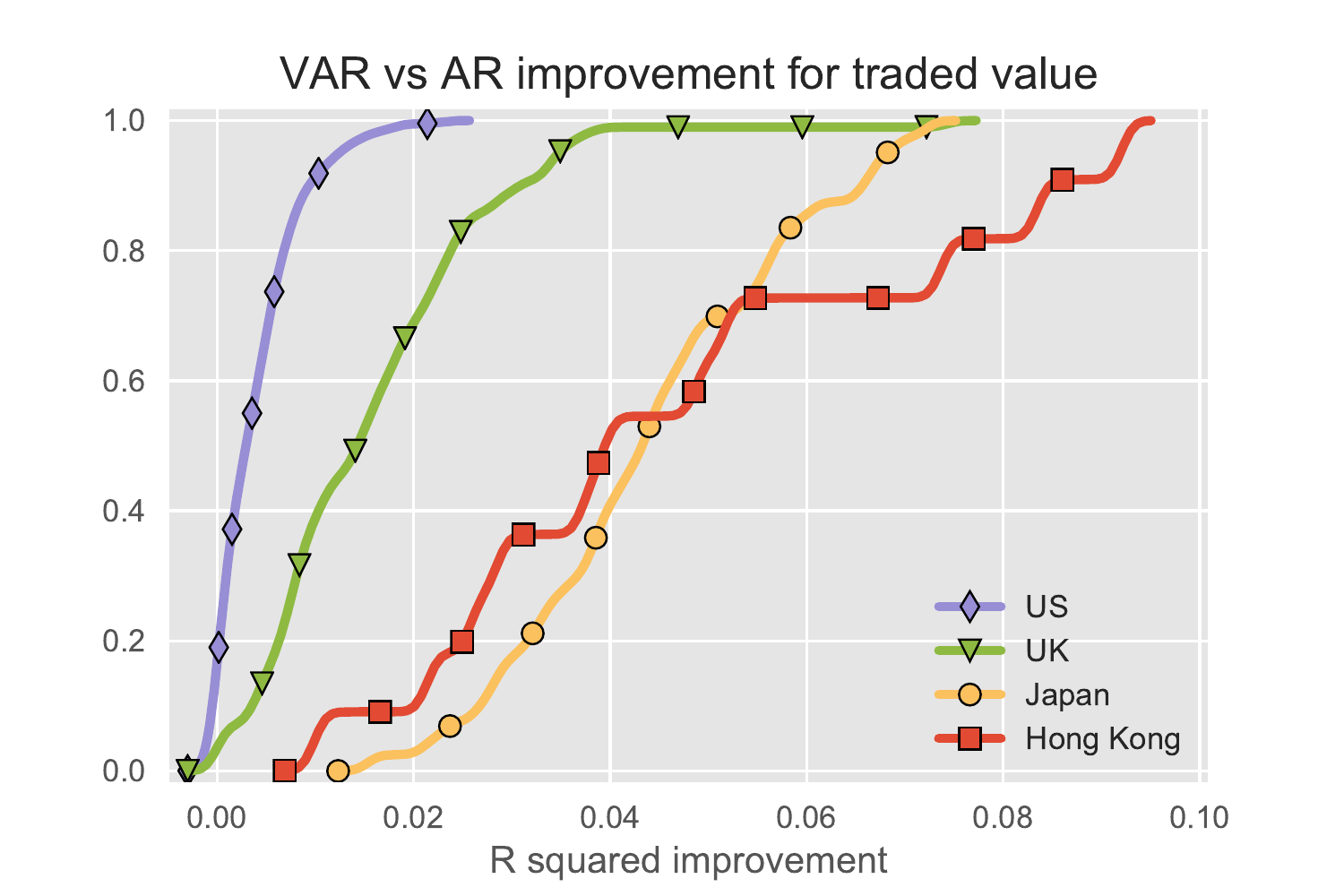}
   \caption[default caption]{Cumulative distribution functions of $R^2$ improvement (difference between VAR's and AR's $R^2$) for traded value. The distribution is estimated separately for each market on the improvements obtained for single stocks from that market. One may notice 'steps' in the distribution for Hong Kong, which occur due to lower number of stocks from this market available for this study.}
   \label{f:improvement1}
\end{minipage}
\end{figure}

Results presented in Table \ref{t:results}, Figure \ref{f:spread_tick} and Figure \ref{f:spread_tick2} legitimize a general observation that there is a country-driven effect: in general, it is easier to predict the considered variables, given their immediate past in the US than in the UK. It is worst for Asian stocks. This \emph{country-driven effect} is difficult to identify clearly; it is probably a mix of effects, including capitalization of listed companies (that are on average greater for the US than for the UK, and then for Japan and Hong-Kong, in general and thus in our database too), and of the tick size. Regulations probably play a role as trading practices too\footnote{Chapter 3 of \cite{citeulike:12047995} underlines trading algorithms using in these countries is different, with more \emph{implementation shortfall} in the US, more \emph{VWAP} in Europe and more \emph{Percentage of Volume} in Asia.}.
 Although the achieved $R^2$ varies by stock, on average the US stocks carry the highest informational content, followed, with few exceptions, by UK, Japanese and Hong Kong stocks. 

The UK market appears to have significantly less structured relation between endogenous information and the average spread in ticks, especially for volatility and bid-ask spread. While the stocks from the other markets form clear clusters, for UK stocks the informational content of the past stretches from very low to fairly high ($R^2$ from 0 to .8), with little or no dependence on the avg. spread in ticks characteristic. This effect is not observed for the book size and was less significant for the traded value.

Another differences between the markets can be observed for traded value. The \emph{smaller} market, the more significant are the differences between univariate and multivariate models. In general, the VAR model helps in lowering the lag\footnote{Note that as we compare one-dimensional and multi-dimensional autoregressive models, the lower lag does not necessarily mean less parameters. Nevertheless, lag is related to the process' \emph{memory} and will be discussed further in the next section.} rather than in attaining higher information content. This however is not the case for traded value of Hong Kong and Japanese stocks, where the difference between AR and VAR is significant in terms of $R^2$.

To conclude, the informational content of the the liquidity variables is dependent on the market. The endogenous information of each variable is generally similar across the stocks traded in the same area (except the UK), which might be related both to the internal regulations and to the market size.

\subsection{Memory effect and causality}
\label{sec:memory_effect}

\begin{figure}[!ht]
   \centering
   \includegraphics[width=\textwidth]{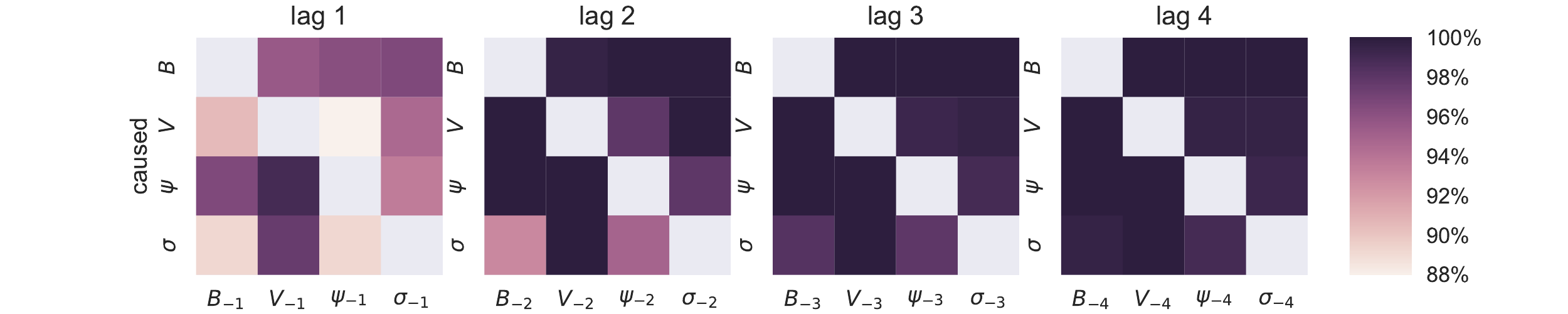}
   \caption[default caption]{Results from Granger $\chi^2$ causality test for the first 4 lags. The shade indicates the proportion of stocks for which the null hypothesis (that there is causality between variables)) was retained under significance level of 0.001. For higher lags (i.e. longer time-span between the potential cause and the caused variable) more than 99\% of tests retained the null hypothesis.}
   \label{f:granger}
\end{figure}

\paragraph{{Causality test}.}
The considered variables expose clear correlation structure between each other (See Figure \ref{cross-correl}). To examine causality between these variables we carried out Granger $\chi^2$ causality tests, separately for each stock, each pair of variables and each lag between them. As presented on Figure \ref{f:granger}, most of the tests retained null hypothesis that there is a causality between given variables, yet whenever the lag, i.e. time difference between the (potential) cause and the dependent variable, was small, in some cases (up to 15\% of test) the null hypothesis was rejected. The higher the lag, the share of rejected tests is smaller (less than 1\% for lag 4, 0\% for lags above 16). 

This suggests that although the variables influence each other, the information between them is not passed instantly. Moreover, the information passing is not symmetric. For instance, the traded value G-causes the next value of bid-ask spread for almost all variables; the opposite is much less common (see Figure \ref{f:granger}). 

The positive results from Granger test reaffirm the legitimacy of multivariate autoregressive modeling. Given these, we discuss the choice of lag and its effects on the results. 

\paragraph{{Choice of lag}.}
The choice of the right number of lags is an important problem in autoregressive modeling. As explained in Section \ref{sec:methodo}, the criterion that we used for the selection of optimal lag was the out-of-sample performance. Theoretically, extending the number of lags should not have the negative impact on the performance. In practice, however, the models with higher numbers of parameters are more prone to overfitting, which might be seen as inference from noise or local features present only in the training set. As a consequence, the lags of the selected models were extended as long as the input series carried more endogenous informational content. 

This approach let us infer how many past observations of the series \emph{cause} the current one, in the sense of Granger, and hence discuss the \emph{'market memory'} of each stock. 

\begin{figure}[!h]
   \centering
   \includegraphics[width =\textwidth]{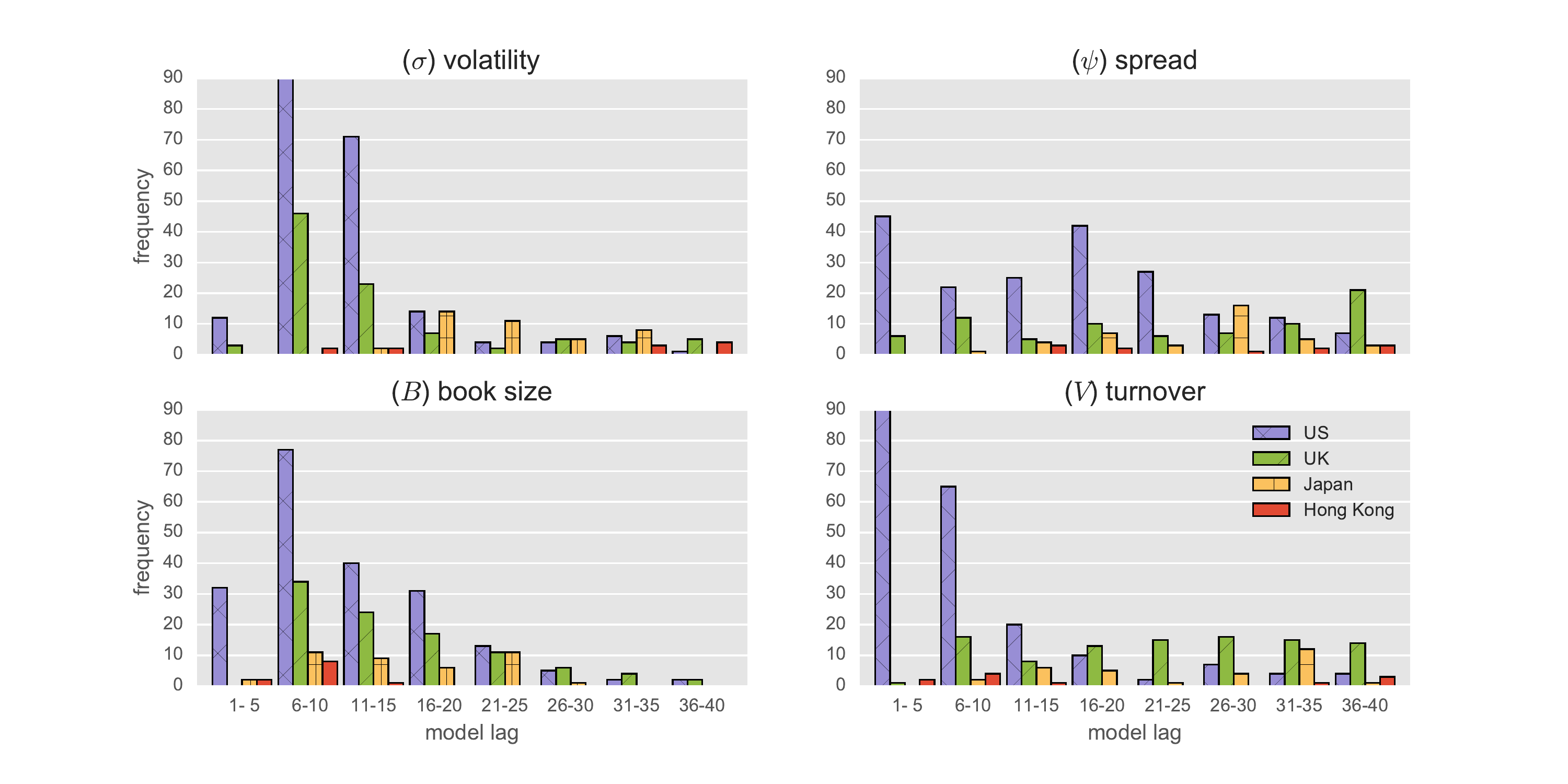}
   \caption[default caption]{Distributions of the most-informative lags for each of the explained variables. }
   \label{f:frequent_lags}
\end{figure}

Figure \ref{f:frequent_lags} presents the distribution of the most informative lags among different countries and target variables. On average, the best models for bid-ask spread had on average significantly higher numbers of lags (18) than the other variables (12-14). These lags correspond to periods of 60 to 90 minutes. However, for certain stocks and features, the numbers of lags were much lower or much higher, including the maximal tested lag (40). Table \ref{t:lags} presents the average lags estimated for each country and variable and compares with the respective results of univariate model.

One may notice that in some cases the results varied by location. For instance, the average lags for American stocks were the lowest, for all the variables except the book size, where Hong Kong stocks had exceptionally small average lag (\~5.6). On the other hand, the average lags for Japanese stocks were often the highest. For volatility, the average lags among Asian stocks were twice as big as for UK and US stocks. 

\begin{table}
 \centering
 \begin{tabular}{lrrrrrrrr}
  \toprule
  location & \multicolumn{2}{c}{Hong Kong} & \multicolumn{2}{c}{Japan} & \multicolumn{2}{c}{UK} & \multicolumn{2}{c}{US} \\    
  \cmidrule(lr){2-3} \cmidrule(lr){4-5} \cmidrule(lr){6-7} \cmidrule(lr){8-9}
  model & AR & VAR & AR & VAR & AR & VAR & AR & VAR \\
  \midrule
  volatility          & 31.5 & 25.5 & 31.7 & 22.6 & 13.9 & 12.8 & 14.9 & 10.7  \\
  bid-ask spread      & 38.3 & 24.4 & 33.6 & 24.9 & 29.4 & 22.7 & 18.2 & 14.9 \\
  book size           & 24.1 &  5.6 & 34.4 & 14.0 & 33.8 & 14.4 & 29.2 & 10.8 \\
  traded value        & 32.7 & 17.6 & 38.2 & 23.2 & 35.1 & 22.4 & 18.4 &  8.0 \\
  \hline
 \end{tabular}
 \caption[default caption]{The average lags of the best AR and VAR models across different variables and markets.}
 \label{t:lags}
\end{table}

We can also observe that in many cases VAR models reduce the lag significantly. In many cases, quite similar $R^2$ was attained by both AR and VAR, yet the latter did not need to look as far back as the former. This effect was particularly apparent for book size for most of the stocks. For the other variables it was observed less often. For instance, since for volatility the univariate models were usually preferable, the VAR model rarely gave any advantage.

\begin{figure}[!ht]
   \centering
   \includegraphics[width =\textwidth]{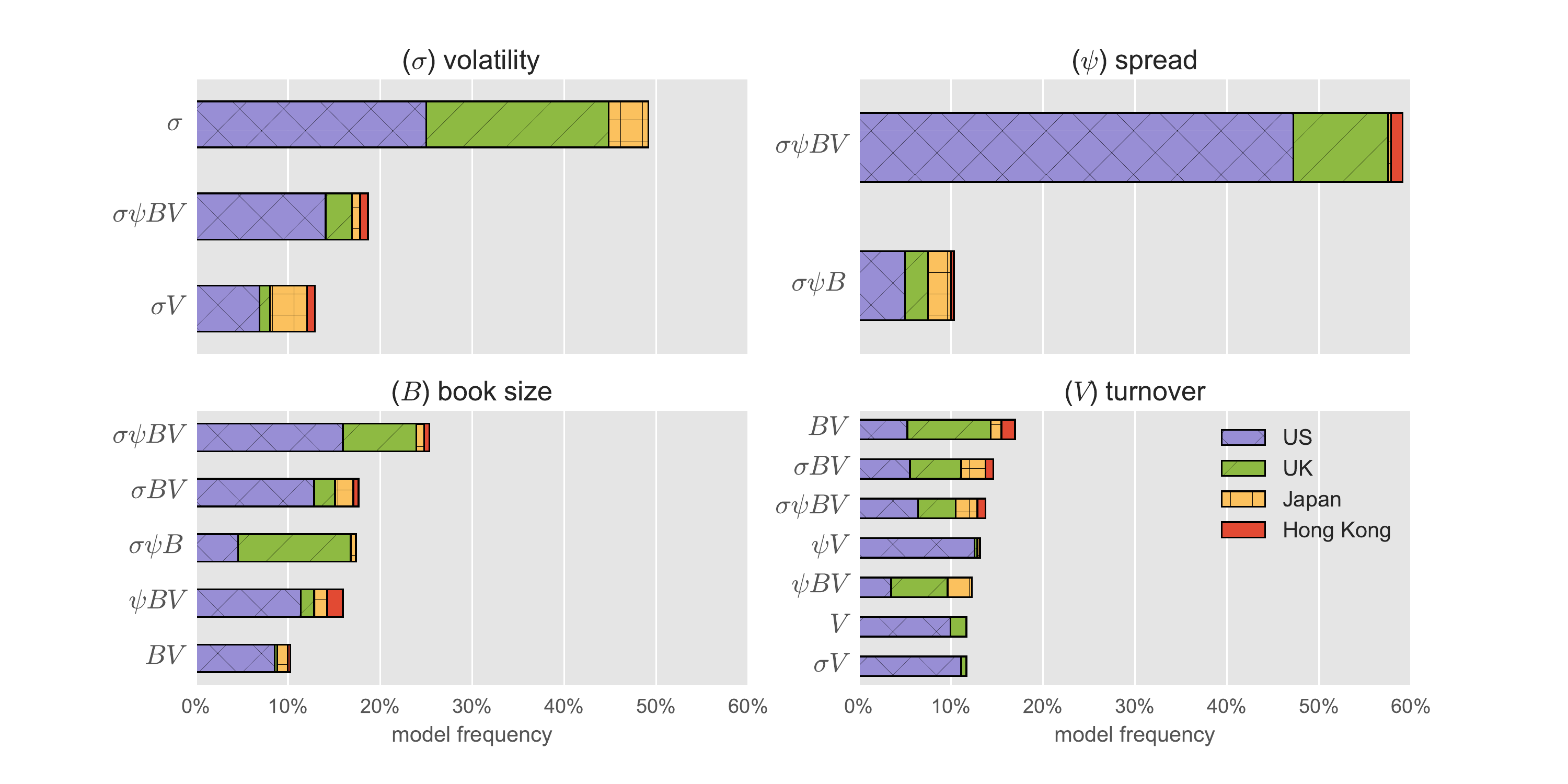}
   \caption[default caption]{Distributions of the most-informative sets of explanatory variables for each of the explained variables. Each of the 16 considered configurations is labeled by the symbols of variables it consisted of, e.g. '$\sigma\psi BV$' denotes the four-dimensional VAR model that included all of the variables: volatility, bid-ask spread, book size and turnover.}
   \label{f:frequent_models}
\end{figure}

\paragraph{{Choice of variables}.}
The variables showed different behavior in terms of preferred VAR model. As mentioned before, for volatility, the most common best model was the univariate one, i.e. the other variables often did not carry additional informational content on volatility. These variables, on the other hand, showed more complicated, diverse relationships between each other. Although four-variable model was often the model of choice (especially for bid-ask spread), for many stocks we observed VAR models of one, two or three variables obtaining the highest out-of-sample $R^2$. Figure \ref{f:frequent_models} summarizes the models that commonly attained the highest informational content of the past among studied stocks. Note that the popularity of each model for a particular variable varied by geographical zone.

Overall, the bid-ask spread has longer memory than the other variables. The informational content of the individual variables can stretch from 15 minutes to over 3 hours of their immediate past, but, except for volatility, the same information can often be found in the lower lags of the other variables.

The plausible explanation for the different behavior of volatility is that the other variables are endogenous features the order book. Although their values also are dependent on the order flow, volatility might be more susceptible to the information exogenous to the microstructure of the market.

\section{Conclusion}
\input{concl.tex}

\bibliographystyle{plainnat} 

\bibliography{small_lehalle.bib}


\end{document}

%% file: intro.tex
The raise of electronic trading not only increased the traded volumes and decreased the bid-ask spread, but also increased the trading frequencies \cite{citeulike:6853976}: market makers are now \emph{High Frequency Market Makers} \cite{citeulike:8423311}
(as an example: Barclays, Goldman Sachs and Bank of America have been replaced in 2016 by Global Trading Systems, Virtu Financial, KCG Holdings and IMC as the only four designated market makers of the New York stock exchange \cite{bloom16MM}),
traders use smart order routers and trading algorithms to seek liquidity and schedule large orders \cite{citeulike:12047995}, while investors reassess more frequently their decisions.
The mix of these different flows shapes liquidity dynamics, and liquidity dynamics drives trading costs and the capability to invest at a predictable price (see \cite{citeulike:12825932,citeulike:13266538,citeulike:13497373} for details about dynamics of the market impact of large orders).

A better understanding of liquidity dynamics is hence of paramount importance for intermediaries, investors and regulators \cite{citeulike:13616374}.
Some academic studies focus on very high frequency dynamics, the most celebrated family being the use of Hawkes processes (see \cite{doi:10.1142/S2382626615500057}, \cite{jaisson:tel-01212087} or \cite{citeulike:12011721}) to explain high frequency price moves (and hence intraday volatility) using the list of past events affecting the orderbook (transactions, cancellations, price moves, etc).
Another high frequency approach is driven by looking at the transitions from one state of the orderbook to another, typically using a Markov chain on queue sizes (see \cite{citeulike:12810809} or \cite{citeulike:8531765}).
Both approaches underline the role of the recent past of liquidity in explaining its future state. These results show that the endogenous component of liquidity dynamics cannot be neglected\footnote{Exogenous shocks affect liquidity dynamics too; \cite{lillo15fx} underlines the added value of news in Hawkes models}. 
Because of the analysis of high frequency data is intensively CPU and memory consuming, these studies usually focus on few typical stocks or futures, having a ``large tick'' (i.e. an average value of the bid-ask spread in ticks being lower than 1.3).

In the context of econometrics, academic paper usually prefer to use ``bins'' of data, typically undersampling price dynamics at a 5 minutes frequency\footnote{The noticeable exception is \cite{bechler2017order}, using bins not in minutes but in traded volume.}. 
A typical metric used in such papers is the \emph{variance ratio} (for instance 30 minutes volatility divided by 5 minutes volatility), as an indirect indicator of the ratio of exogenous over endogenous part of price dynamics (see \cite{asbrouck95} or \cite{nguenang2016evaluating} and references herein). 
Some papers build ad hoc models, like 
\cite{lefo08} that combines a PCA on previously traded volume on forty French liquid stocks (the components of the CAC 40 index in September 2014) and an ARMA model to predict the volume traded during the next five minutes. Another example is \cite{citeulike:13562625}, in which authors focus on exogenous shocks on liquidity.
All these papers implicitly assume autocorrelation (or not) of their variables of interest, without providing stable reference statistics over a large universe of stocks, covering more than one geographical zone.

Another important field relying on five-minute to one-hour ``bins'' of liquidity characteristics is optimal trading. Most optimal trading (or optimal liquidation) algorithms used by market participants focus at this kind of time scale \cite[Chapter 3]{citeulike:12047995}.
Very few papers in optimal trading include autocorrelations in their theoretical framework, simply because of the mathematical complexity generated by path-dependence for stochastic control. Correlations between different liquidity-related variables have been used (see \cite{LEHW08}) and some path dependencies have been recently introduced (see \cite{citeulike:13125577} for Hawkes process, \cite{citeulike:14448873} for Ornstein-Uhlenbeck information on prices, and \cite{citeulike:13587586} to account for flow driven autocorrelations).

Last but not least, practitioners use TCA (Transaction Cost Analysis, see \cite{tabbTCA08} and \cite{kissell2004practical}) to assess the quality of the execution their brokers provided, and the efficiency of their own dealing desks. Very often the TCA reports are made of plain averages, which can be completely misleading in presence of autocorrelations. Primarily, TCAs aim to isolate the adequacy of some trading choices, \emph{given the expected state of liquidity on the given day}. Ignoring autocorrelations distorts the reference expectations.

This paper is dedicated to establish benchmark models on intraday liquidity dynamics at this intermediate time scale (few times 5 minutes), so that specific studies or TCA using this nature of data can use our findings as a worldwide compass.


Our basic methodology is to 
systematically fit linear models on time series of liquidity related variables; and to exhibit and comment on the obtained results.
Our goal is on the one hand to document the liquidity dynamics at this intermediate scale to provide a benchmark for further analyses 
and on the other hand to compare the results between stocks and across geographical zones.


By ``\emph{endogenous dynamics}'', we mean that our models only use the past values of the explained variables as regressors: they do not use any database of external market events or fundamental news dataset on the considered stocks.
By ``\emph{linear dynamics}'', we mean we only use vectorial or scalar standard autoregressive models \cite{citeulike:7342108}.

Thanks to Wold theorem, it is well known any centered weakly stationary process can be decomposed into a regular part and a singular part, and that in non-degenerate cases they can be fit on the right and left terms of an ARMA (AutoRegressive Moving Average) process \cite{citeulike:7342108}.
We restrict our analysis on ARMA$(p,1)$ models, meaning we will only allow a white noise on the right side of the ARMA.
This choice is driven by the fact it is very difficult to compare an ARMA$(p_1,q_1)$ and an ARMA$(p_2,q_2)$, but easier to compare an ARMA$(p_1,1)$ and an ARMA$(p_2,1)$.
As a consequence, the reader should keep in mind that some of the obtained models could be replaced by full ARMA ones. When it is possible to fit a full ARMA instead of an AR, the first parameter $p$ of the obtained full ARMA will probably be a little smaller than ours, at the cost of replacing our white noise by a colored one\footnote{An ARMA$(p,q)$ and an ARMA$(p',1)$ both properly fitted on the same empirical time series will usually verify $p\leq p'$ (see for instance \cite{geist2011kalman} for an illustration of switching from AR modeling to full ARMA modeling ).}.

Another viewpoint on our approach would be to consider \cmt{our models} as ``\emph{linear filters}'' on liquidity variables rather than the best possible models of their dynamics. In this spirit of filtering we will use \emph{Cross Validation} to choose the number of lags of our AR or VAR models.
This technique is widely used in the statistical learning community to choose the number of parameters (i.e. ``hyperparameters'') of a model (like the number of units of a neural network) \cite{cor16cross}. 

Our liquidity-driven variables of interest are the following quantities, estimated on consecutive bins of five minutes from the opening to the closing of trading sessions:
\begin{itemize}
\item the traded value (or ``turnover''), i.e. is the value (in the local currency) of the shares traded; we use the turnover instead of the number of shares to be robust to split or reverse splits,
\item the average bid-ask spread (sampled just before each transaction in the considered bin),
\item the ``volume on the book'' (average of the best ask volume and the best bid volume just before each trade in the bin),
\item the Garman-Klass estimate of the volatility\footnote{cf. \cite{GAR80}; this estimate is robust and very close in practice to more sophisticated ones, like \cite{scales05} or \cite{robert2010new}.} that uses the open, high, low and close prices of the bin.
\end{itemize}
We use the data aggregated from trading venues hosting more than 1\% of the traded value on each considered stock.
This means we mix trades done on Nyse, BATS and NASAQ for the same stocks in the US, and we use the consolidated tape\footnote{When no consolidated tape exists, like in Europe, we consolidate the bid and ask of trading venues ourselves, using timestamps provided by markets.} to compute the bid-ask spreads.
Our database of 300 stocks covers UK, US and Asian stocks and spans five years from 2011 to 2016.

We systematically fit AR models on each of them (i.e. predicting the next value of the variable using its own past) and VAR models on each of their combinations (i.e. predicting the future of each variable using the past of a subset of all variables). 
The contributions of this paper include several \emph{stylized facts} that show differences as well as interdependencies between studied variables across different markets. 
Firstly, large-tick stocks have more predictable book size and less predictable spread, volatility and turnover, than the small-tick stocks, given their immediate past; we call this a \emph{tick size effect}.
Secondly, volatility and turnover are also subject to the \emph{liquidity effect}: the higher the market capitalization of a stock in a considered market, the shorter the memory in these features. 

Thirdly, there exist differences that could be named country-driven in the structures of the markets, that we call a \emph{liquidity effect}. US stocks carry more endogenous information than stocks from the other markets and have shorter memory. UK stocks have varying informational content, independent on individual liquidity and tick size. Asian stocks, on their side, carry less endogenous information and have longer memory. We focus on a representation of such characteristics (amount of endogenous information and memory) with respect to the relative capitalization of the considered stocks inside each country. This representation succeeds in explaining the amount of endogenous information and the length of the memory for the volatility, the traded volume and the book size.
Lastly, we observe the \emph{memory effect}. Volatility has shorter memory and often carries whole informational content of all considered variables, i.e. information on the other variables' past does not improve prediction of volatility. In terms of Granger causality\footnote{It is not the first time Granger causality is studied on five-minute intervals. For instance \cite{dar03idca} tested it on the joined volume - volatility dynamics and found some on the US market.} (see \cite{gran88caus}), volatility and turnover are less often caused by the the other variables' immediate past, than the opposite. 

The paper is structured as follows:
we start by detailing the variables, the databases and the pre-processing (e.g. stationarizing the time series) in Section \ref{sec:methodo}.
Section \ref{sec:illustr} provides detailed results on few well-chosen stocks and comments them. 
We conclude in Section \ref{sec:results} by the presentation of empirical findings on the whole dataset, and put the emphasis on few stylized facts: the liquidity effect, the tick-size effect, and a country-driven effect. We end this last section by the study of what we call a memory effect: using VAR in place of AR models reduces the needed lags to achieve the same level of explanation.


%% file: stationarize.tex
\begin{figure}
   \rule{9em}{0pt} AstraZeneca (UK) \hfill 3M \hfill~\\
   \includegraphics[width = .48\linewidth]{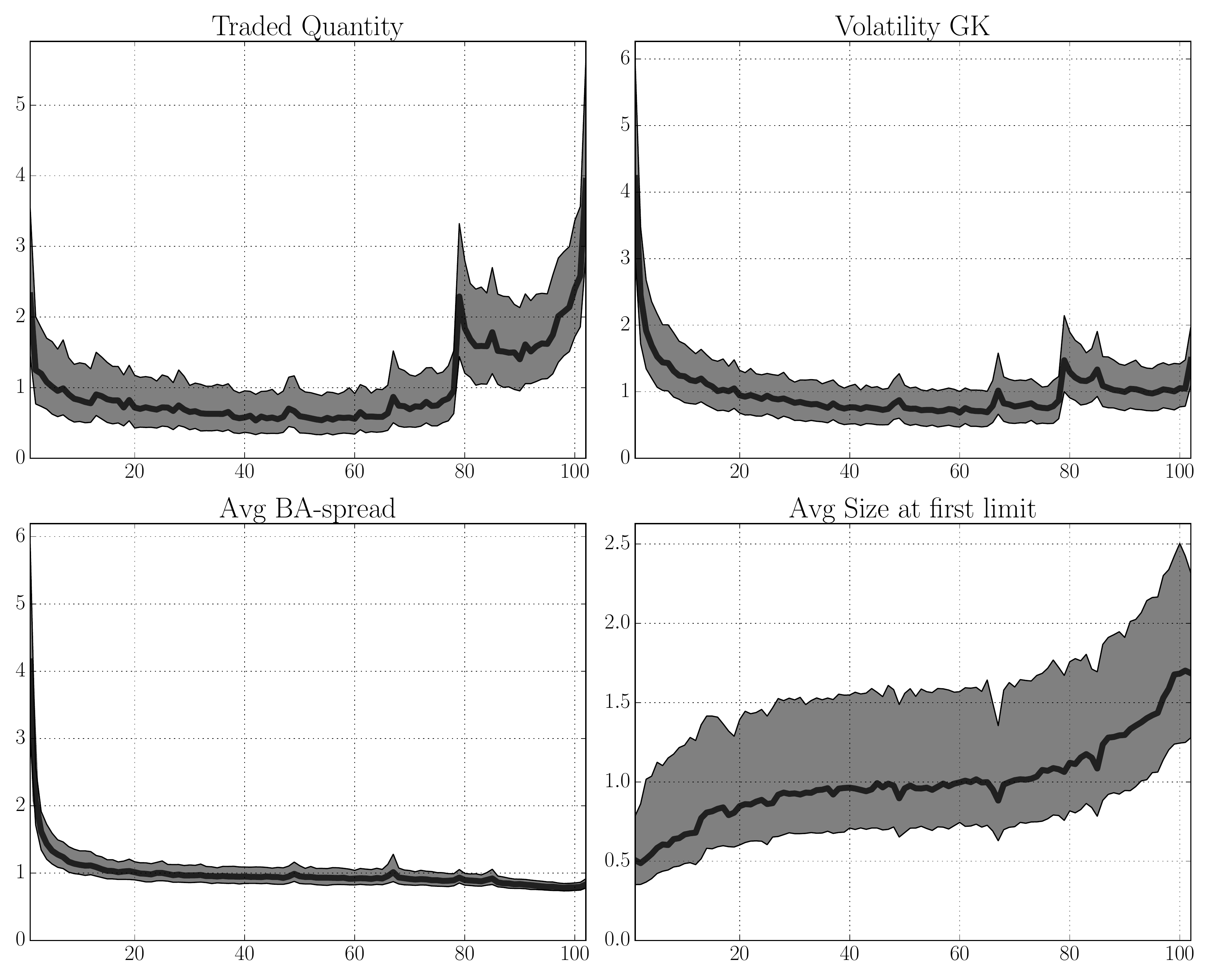}
   \includegraphics[width = .48\linewidth]{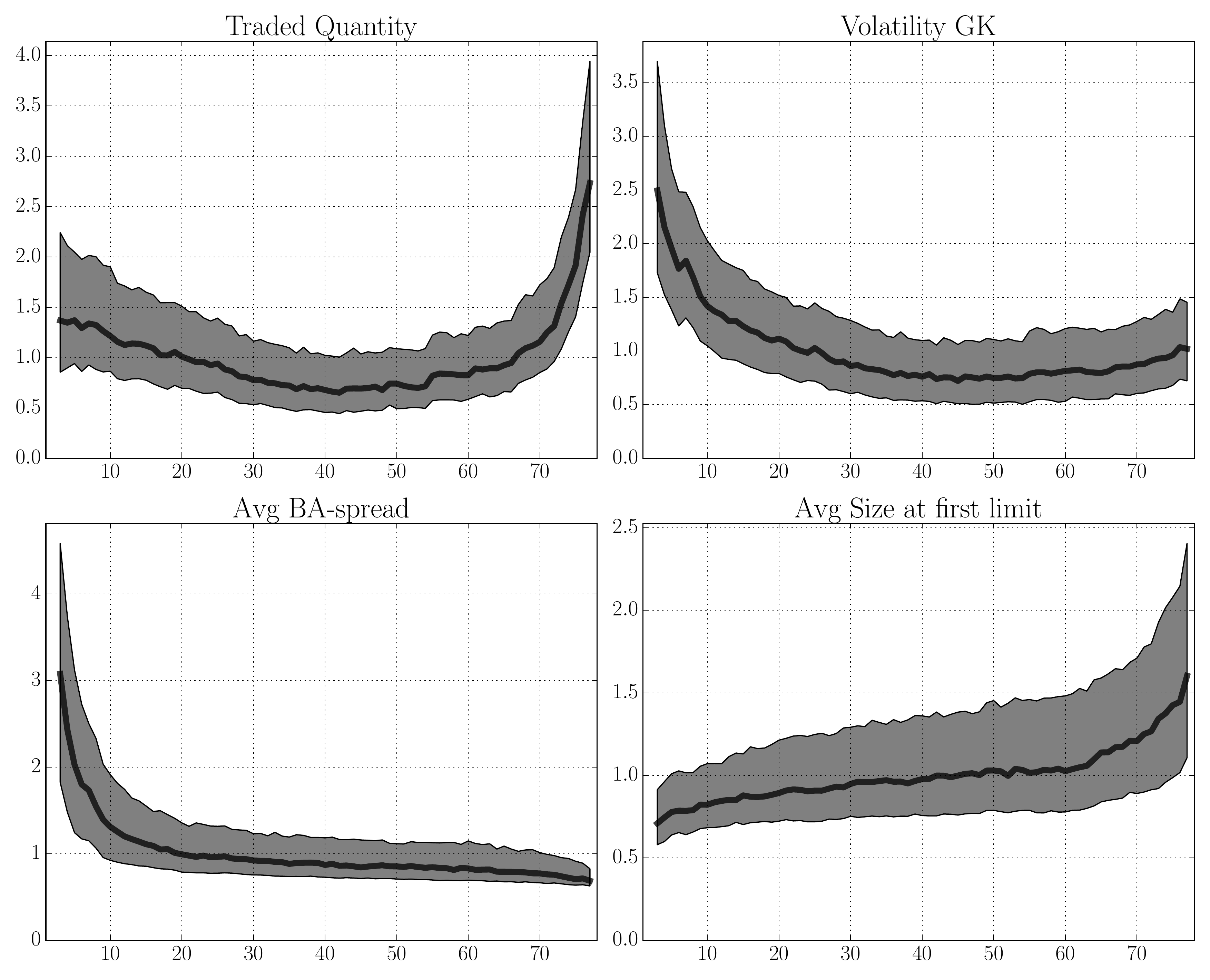}\\
   \rule{9em}{0pt} FUJIFILM (UK) \hfill Cathay Pacific Airways \hfill~\\
   \includegraphics[width =.48\linewidth]{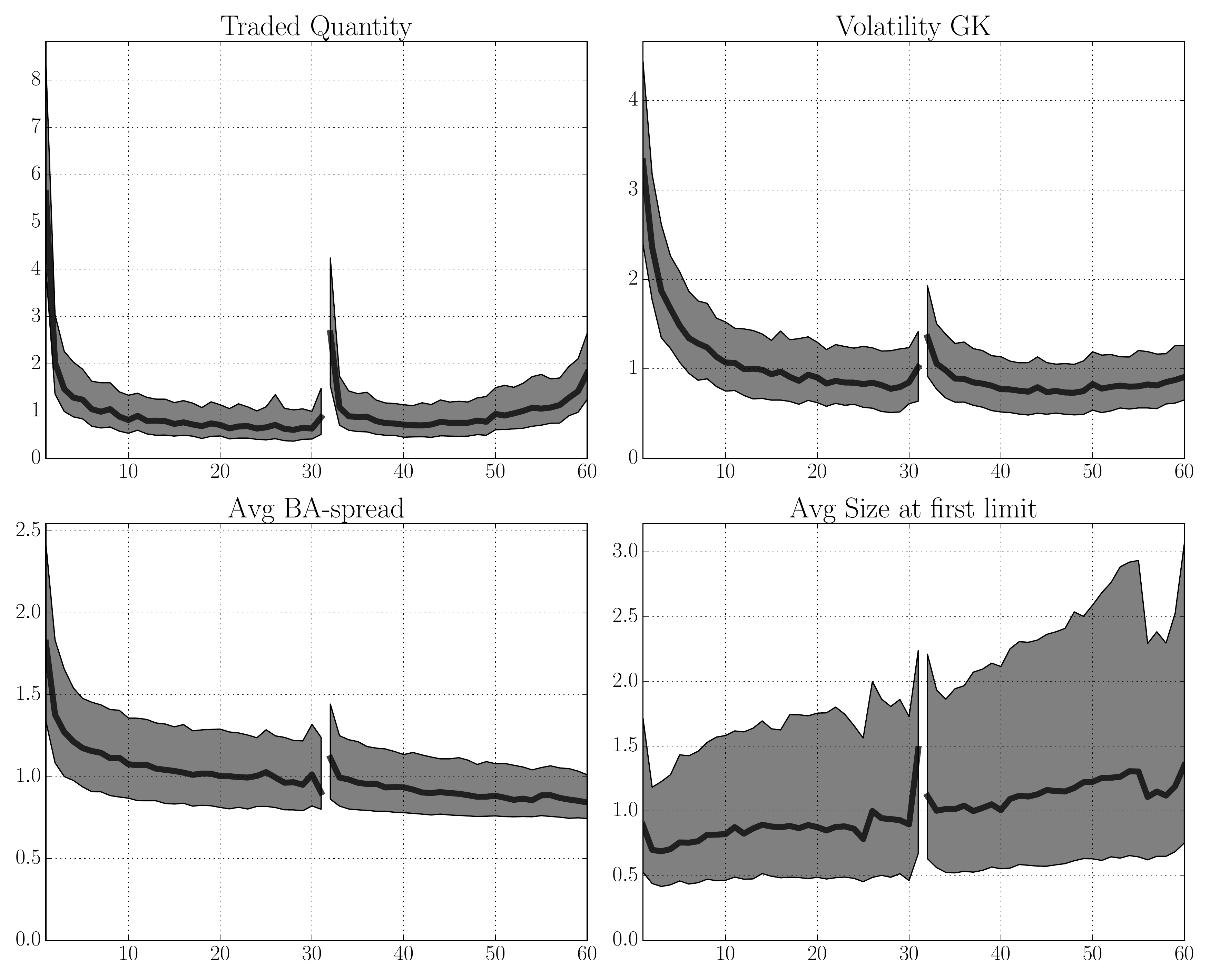}
   \includegraphics[width = .48\linewidth]{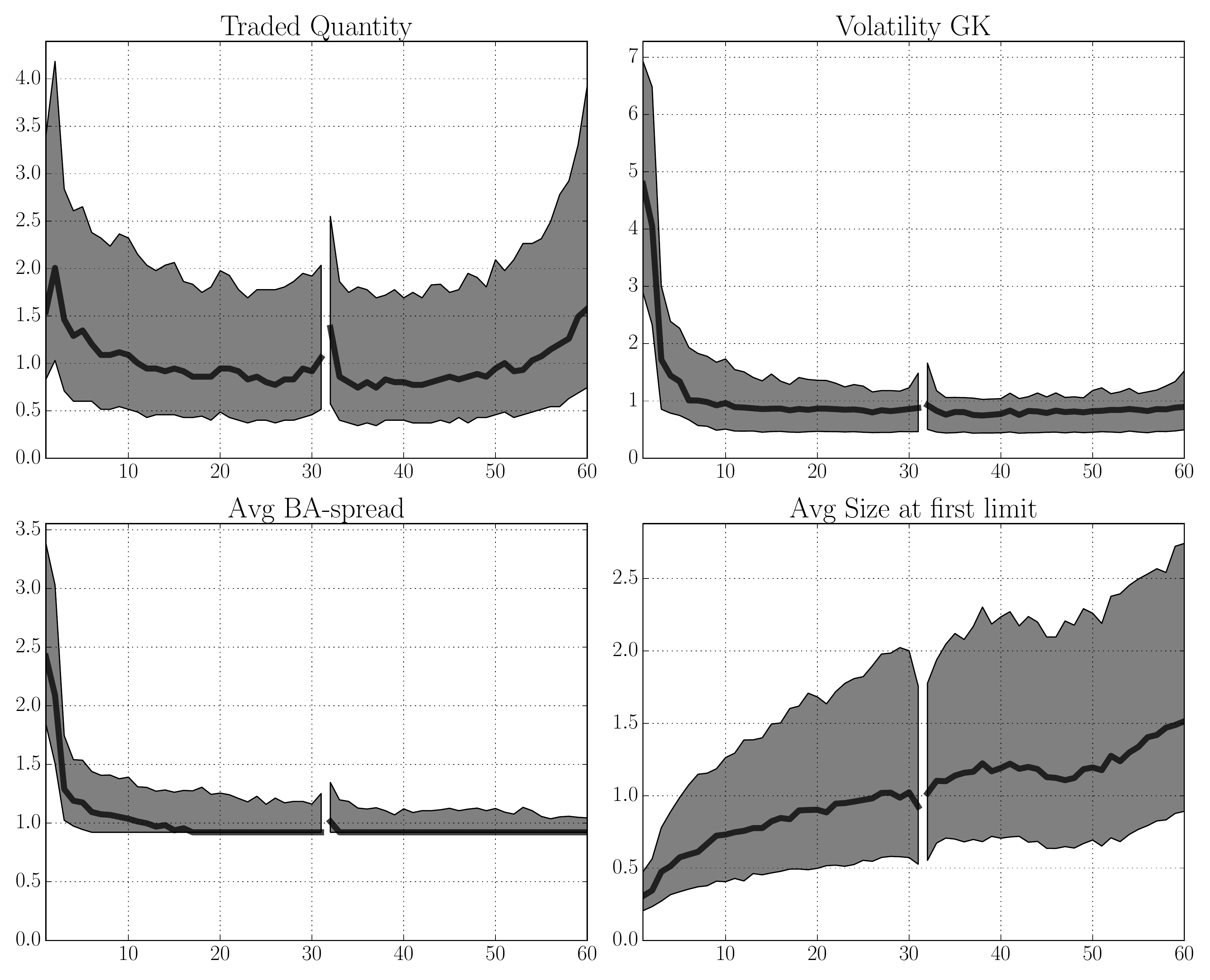}\\
   \caption{Median, 25\% and 75\% quantiles for our four liquidity variables on AstraZenaca (UK stock), 3M (US Stock), FUJIFILM (Japanese Stock), Cathay Pacific Airways (Hong Kong Listed stock). The vertical units are chosen such that the average value of the day is 1; on the horizontal axis we use the ``bin number'' (the consecutive number of non empty 5 minutes bins); meaning during the Tokyo or Hong Kong lunch break we do not count any ``bin'', but we use a blank vertical line to position the lunch break.}
   \label{fig:AZN:curves}
\end{figure}

\paragraph{Stationarization.}
It is well known liquidity variables have an intraday seasonality (see for instance \cite{citeulike:12047995}, or Figure \ref{fig:AZN:curves}): the volatility and the traded volume are higher immediately after the opening, and quite flat during the mid-day. At the end of the day the traded volumes increase, forming a ``U-shape''; the volatility too, but less sharply. 

For European stocks the opening of US market triggers a jump in volatility and volume. The disclosure of macro-economic news on US markets (like the \emph{non farm payrolls}), one hour before the openning of US markets, impacts traded volumes and volatility too. 

Fundamental reasons support a relation between the bid-ask spread and the volatility: market participants acting as market makers (or at least \emph{providing liquidity}, i.e. contributing to the bid-ask spread) fear market risk (see \cite{RePEc:eee:jfinec:v:9:y:1981:i:1:p:47-73} for an economic explanation and \cite{citeulike:9304794} for an applied mathematical one); thus theory predicts the bid-ask spread is large when the volatility is high. It is not that true before the trading closes, since liquidity providers targeting a flat inventory at the end of the day have an incentive to tighten the bid-ask spread, attempting to obtain the fee rebate for limit orders, instead of consuming liquidity via marketable orders, paying fees.

The bid-ask spread is moreover bounded by the \emph{tick size} (i.e. the minimum allowed price increase, see \cite{citeulike:13800065} for a discussion about the influence of the tick size on the bid-ask spread). Because of the impossibility for liquidity providers to post limit orders inside the bid-ask spread (and thus tighten it) when the spread value is one tick, they tend to post more quantity at existing first limits (on bid and ask sides). As a consequence the average size on first limits (i.e. our \emph{book size} variable) increases. The book size and the bid-ask spread hence usually vary an opposite way: when the bid-ask spread is large the book size is small, and the reverse.

\begin{figure}[h]
   \centering
   \includegraphics[width =\textwidth]{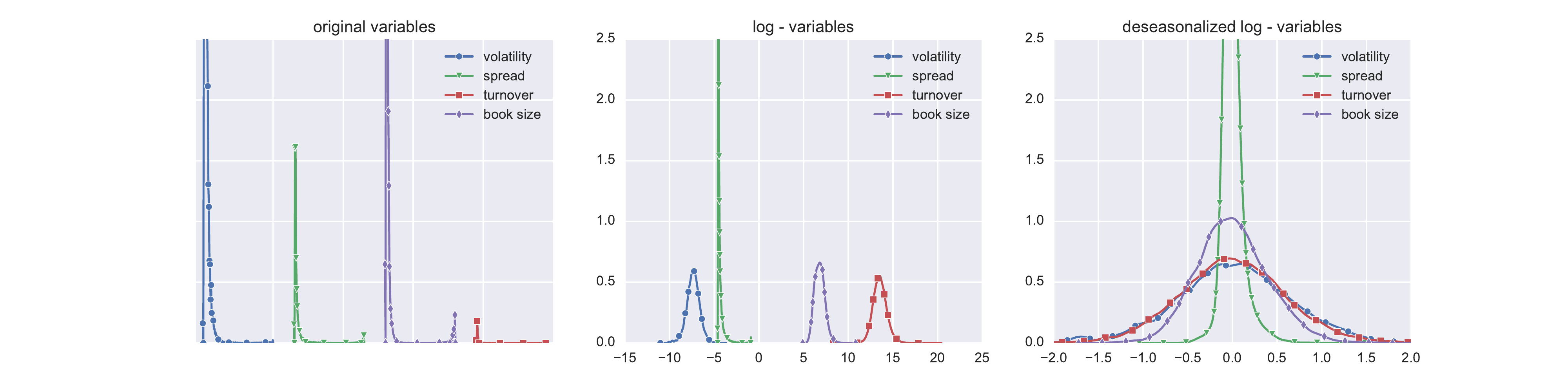}
   \caption[default caption]{Estimated density functions of the distributions of the original variables (left), after taking logarithms (middle) and after deseasonalization (right) for Dahaner Corp. The original variables have highly skewed distributions and take values in very different orders of magnitude, hence the densities in the left plot are not-to-scale.}
   \label{f:normalization_densities}
\end{figure}

These effects introduce typical intraday seasonalities; it is needed to remove them to focus on dynamics \emph{around this average behaviour}.
Since all variables are positive, we take their logarithm to symmetrize them and then we remove the seasonality. Using the notation $x(d,\tau)$ for the value of variable $x$ on day $d$ at bin $\tau$ (i.e. hour is $\tau_0 + \delta \tau\cdot \tau$, where $\tau_0$ is the opening hour and $\delta \tau$ is our bin size --5 minutes--), the pre processing is then
$$y(d,\tau):=f(x(d,\tau)) = \log x(d,\tau) - \underbrace{\frac{1}{D}\sum_{d'=1}^D\log x(d',\tau)}_{\log {\bar x}(\tau)},$$
computed over all the available days $D$ on our dataset.
Figure \ref{f:normalization_densities} shows the effects of the stationarization procedure on Dahaner Corp. stock (US traded stock).

%% file: liquidity.tex
\subsection{Liquidity effect}
\label{sec:likeffect}

\begin{figure}[h!]
   \begin{subfigure}{.5\textwidth}
   \centering
   \includegraphics[width=.8\linewidth]{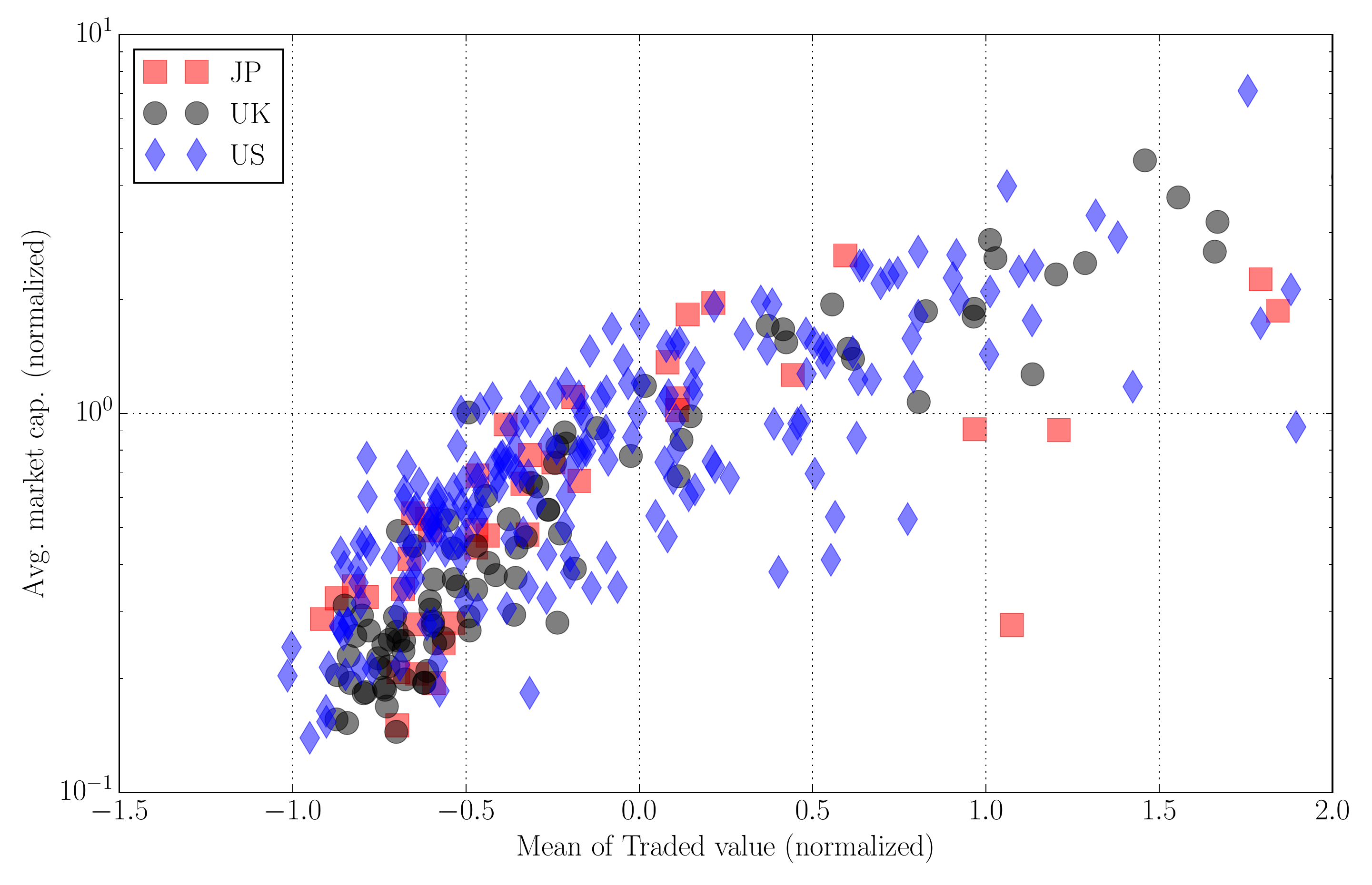}
   \end{subfigure}
   \begin{subfigure}{.5\textwidth}
   \centering
   \includegraphics[width=.8\linewidth]{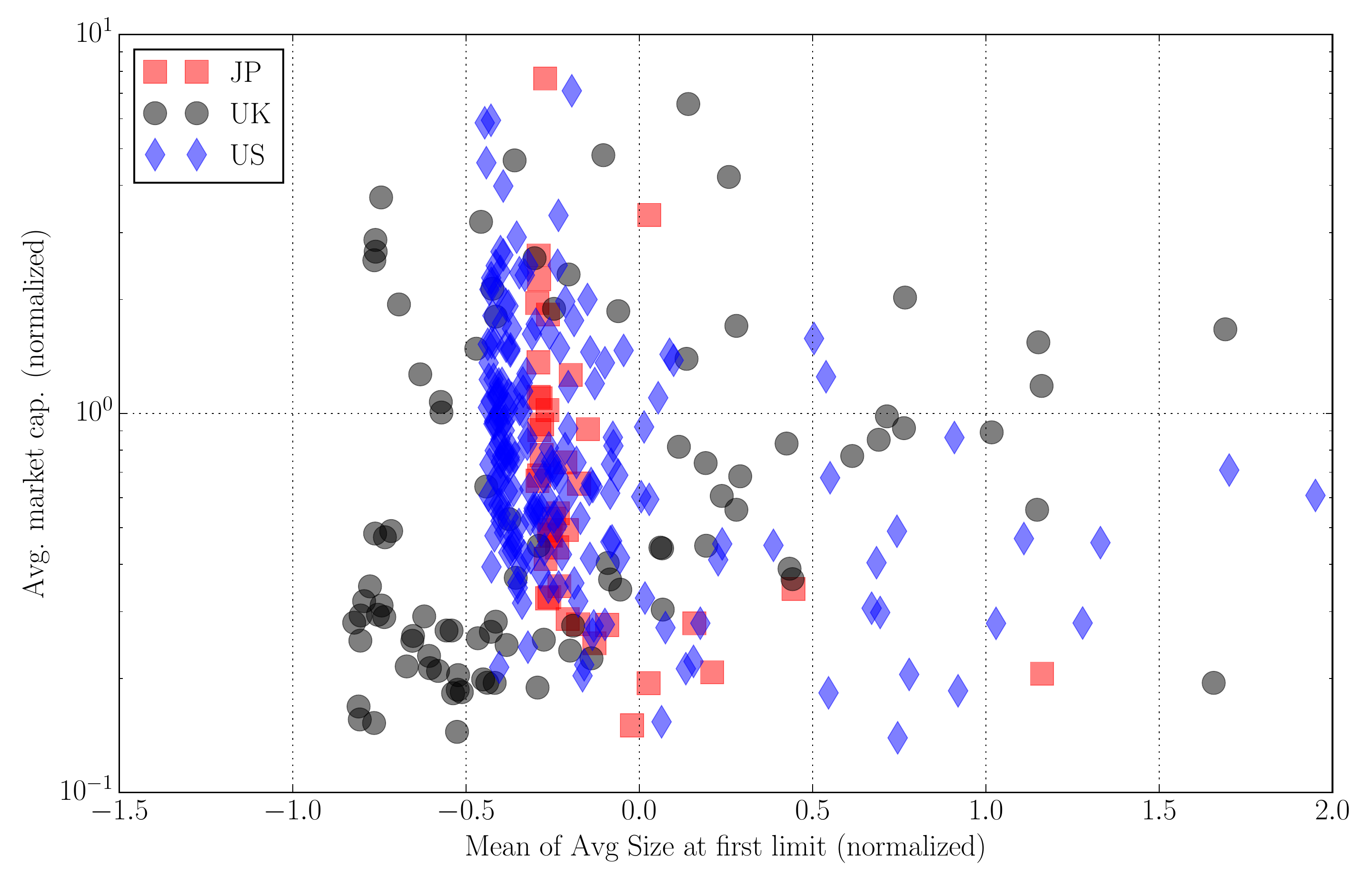}
   \end{subfigure}
   \caption{Average value of traded value (left) and book size (right) compared to the Market Capitalization (MC) of the stock. MC are normalized a cross-sectional way, on each zone. One can see the average traded value is clearly linked to the capitalization.}
   \label{fig:desc:liquidity}
\end{figure}

\paragraph{Towards a better definition of liquidity.}
Up to now we used our four variables (traded value, volatility, Bid-Ask spread and volume at first limits) as a proxy to the ``market liquidity'': the liquidity available to trade for investors on electronic markets during the continuous auctions.
On each of them, we built dynamical models made of one intraday seasonality times a proportional auto-regressive part:
$$x(d,\tau) = {\bar x}(\tau) \cdot \prod_k \left(\frac{x(d,\tau-k)}{{\bar x}(\tau-k)}\right)^{A_k} \cdot e^{\epsilon(d,\tau)} .$$

Practitioners often to measure the liquidity of a stock as its free float market capitalization: since this market capitalization is the amount of dollars that can be traded (on the paper), it is natural to believe the more of such ``tradable dollars'', the easier of an investor to buy or sell a position, and hence the more ``\emph{liquid}'' the stock.
Figure \ref{fig:desc:liquidity} shows the relationship between our four variables and the average market capitalization of our Japanese, British and American stocks from June 2011 to March 2016. The average value $\langle x\rangle$ of the variable is computed during the whole period for each stock $k$ of zone\footnote{\emph{zone} is Amerian, Japan or UK.} $z$, it is then cross-renormalized: ${\langle x(k,z)\rangle}$ is replaced by
$$R({\langle x(k,z)\rangle}) = \frac{{\langle x(k,z)\rangle} - \mbox{mean}_{k'}{\langle x(k',z)\rangle}}{\mbox{std}_{k'}{\langle x(k',z)\rangle}},$$
where the mean and std are computed over all the stocks of the same zone. The market capitalization $C(k,z)$ is cross-renormalized too, dividing it by its standard deviation of all the stocks of the same zone.
One can clearly see the relation between the average traded value in five minutes and the capitalization of the stock. The relation is similar across zones. For the other variables, there is no relationship between their average value during five minutes and the capitalization of the traded stock.


\begin{figure}[h!]
   \begin{subfigure}{.5\textwidth}
   \centering
   \includegraphics[width =.8\linewidth]{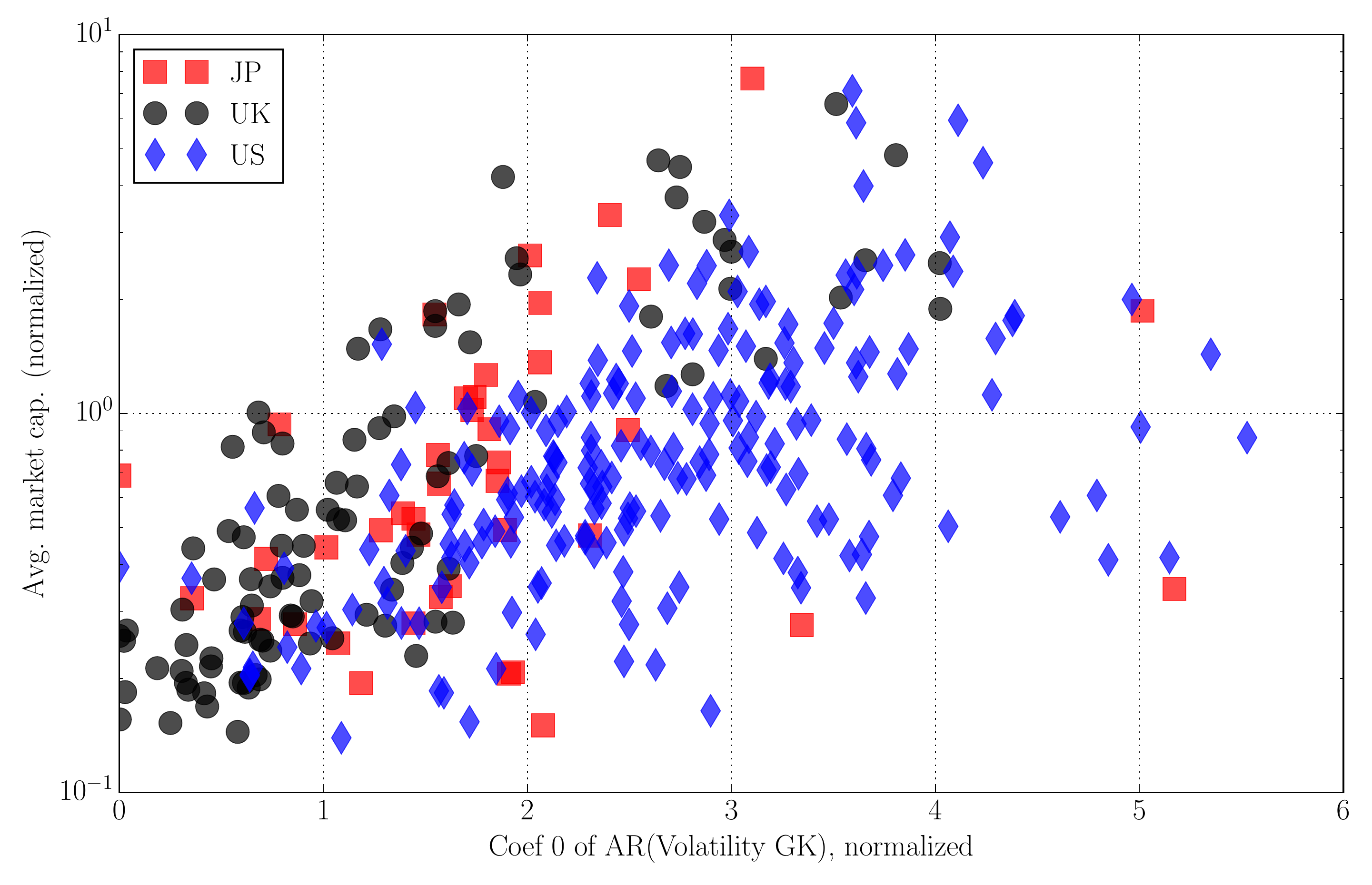}
   \end{subfigure}
   \begin{subfigure}{.5\textwidth}
   \centering
   \includegraphics[width =.8\linewidth]{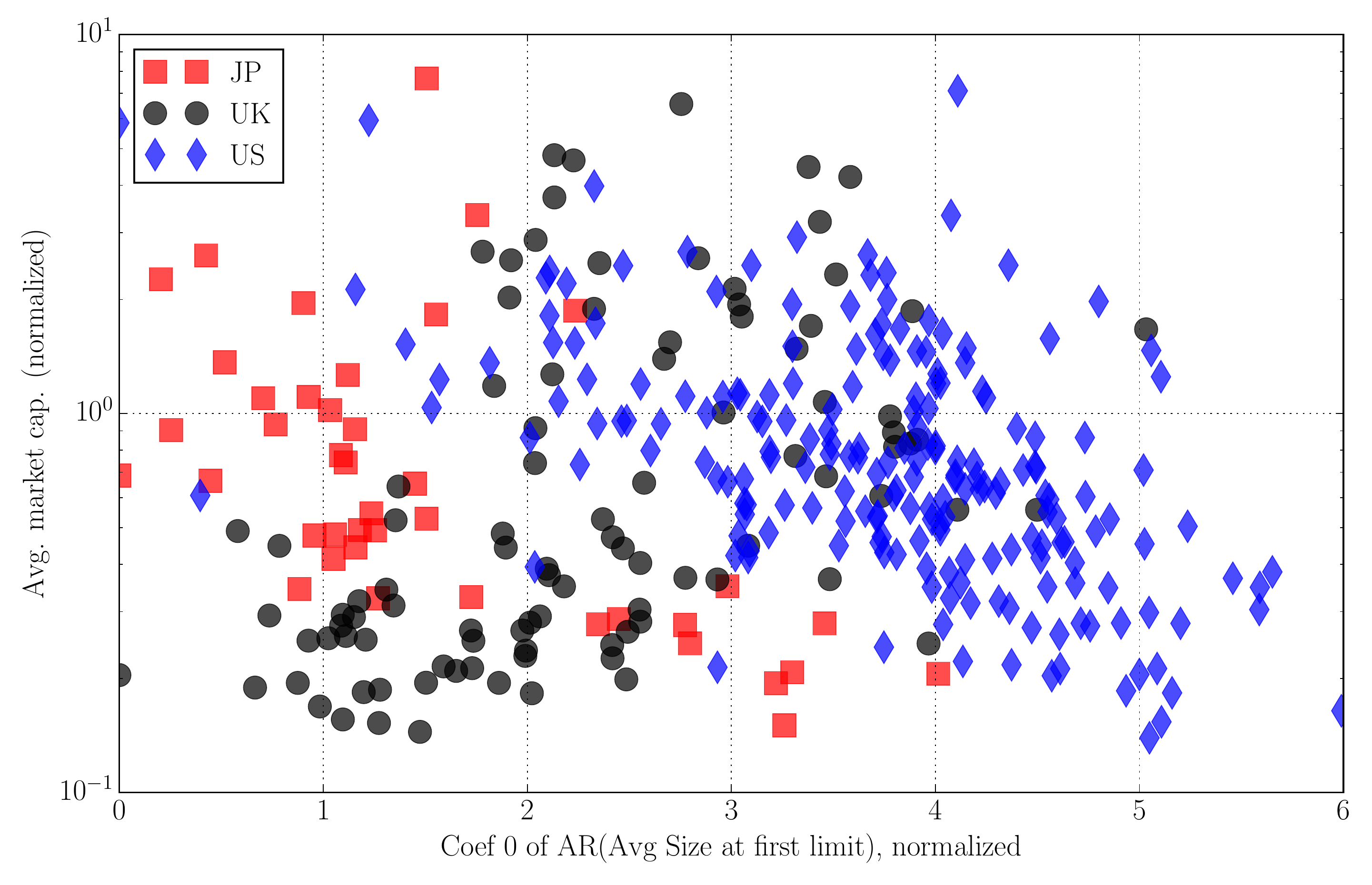}
   \end{subfigure}
   \caption{First value of an AR models ($x$-axis) for volatility (left) and book size (right) compared to the Market Capitalization (MC) of the stock. MC are normalized a cross-sectional way, on each zone.}
   \label{fig:A0:liquidity}
\end{figure}

\paragraph{Memory length of intraday short term variations.}
One of our findings is that the modeling of oscillations around the intraday seasonality exhibits a dependence to market capitalization.
Figure \ref{fig:A0:liquidity} shows how the market capitalization of a stock as a function of the first coefficient $a_0$ of an AR model\footnote{with the following notation $\sum_{i=0}^P a_i X_{t-i} = \varepsilon(t)$.} fitted on our variables dynamics at five minutes (once deseasonalized). 

Note $\lambda$ can be read as ``the larger $\lambda$, the shorter memory''. Our findings are thence: the more liquid in the capitalization sense, the shorter term memory of volatility and traded value around the intraday seasonality; and the more liquid in the capitalization sense, the less dependence to past values in amplitude too.

\paragraph{Summary of the ``Liquidity Effect''.}
These findings are compatible with the following hypothesis: if a liquid stock sees its volatility or traded value going away from its ``usual behavior'', the faster it will come back to normal. 
It is weakly true for the Bid-Ask spread (probably the discrete and bounded nature of the spread forces its dynamics to be similar for most stocks, somehow independently from the market capitalization of the stock). The tick size is certainly of paramount importance in shaping the short term dynamics of the Bid-Ask spread (see the ``\emph{tick effect}'' documented in the next Subsection). Nor the dynamics of volume on first limits or even its average value is disconnected from the market capitalization of the traded stock.

\begin{figure}[h!]
   \begin{subfigure}{.5\textwidth}
   \centering
   \includegraphics[width =.8\linewidth]{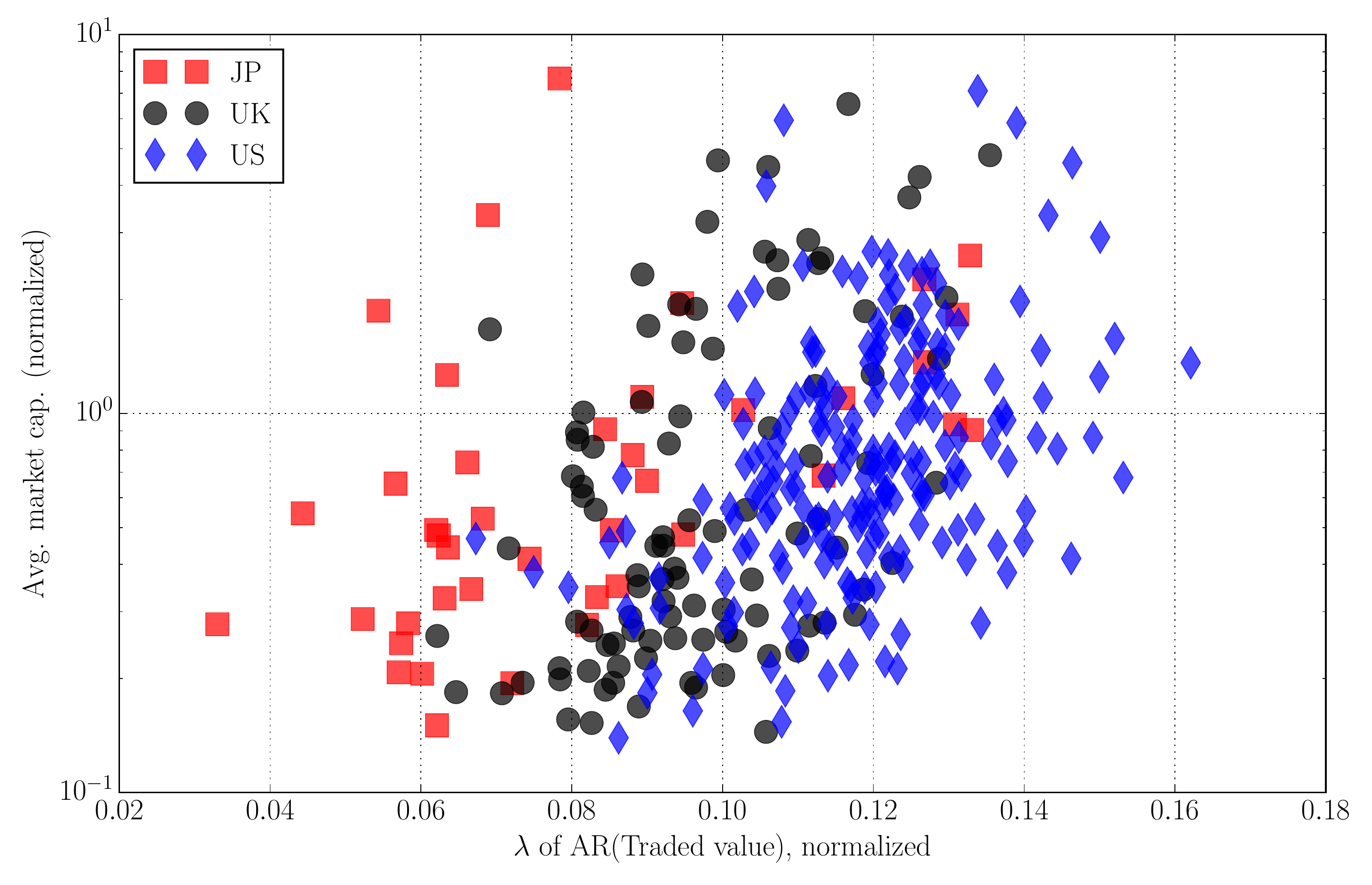}
   \end{subfigure}
   \begin{subfigure}{.5\textwidth}
   \centering
   \includegraphics[width =.8\linewidth]{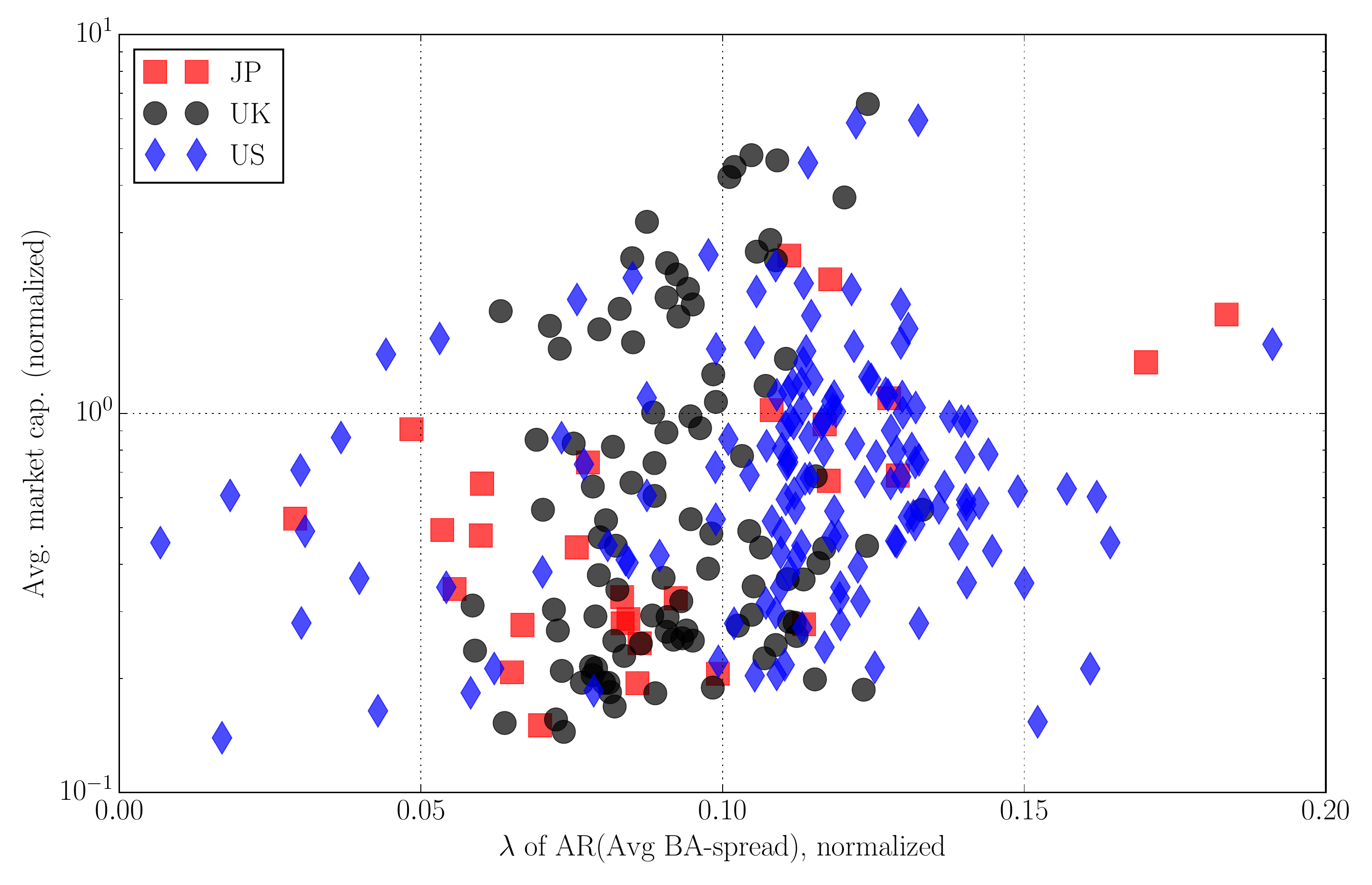}
   \end{subfigure}
   \caption{Memory term $\lambda$ fitted on an AR model ($x$-axis) for traded value (left) and avg. bid-ask spread (right) compared to the Market Capitalization (MC) of the stock. MC are normalized a cross-sectional way, on each zone.}
   \label{fig:lambda:liquidity}
\end{figure}


%% file: concl.tex
In this paper we present a systematic study of the endogenous dynamics of four liquidity variables at the 5 minutes time scales. Once we removed the intraday seasonality of the logarithms of the traded value, the (Garman-Klass) volatility, the bid-ask spread and the size of the book (i.e. the average size at first limits), we model their dynamics a linear way (i.e. via AR or VAR models) using their own past and the past of other variables. 
Hence our models focus on the multiplicative variations around the seasonality of each variable. 
We used cross validation and took the out of sample $R^2$ as our main criterion to draw conclusion from this unique empirical study covering 300 worldwide stocks over five years. 

We isolate some stylized facts:
\begin{itemize}
\item A \emph{liquidity effect}: The larger the market capitalization of a stock, the smaller the length of the past influencing volatility and traded value.
\item The informational content of the book size dynamics deeply relies on a ``\emph{tick size effect}'', for all zones, all stocks: the smaller the bid-ask spread in ticks, the easier to predict the book size.
\item A \emph{country-driven effect}: For the three other variables (volatility, traded value and bid-ask spread): the length of the past needed to anticipate current value is far smaller for US stocks (around one hour) than for stock from other zones (around two and a half hour for UK and Asia). 
Moreover, the informational content of the variables on their endogenous dynamics is higher on the US than on other zones.
\end{itemize}
The specificity of US stocks dynamics is difficult to explain. 
One explanation could be the US market is faster than others, in terms of liquidity. But having an out of sample $R^2$ twice higher in the US than on other zones (i.e. it is easier to anticipate the future values of volatility, bid-ask spread and turnover using their own past only) means more than just a speed difference.
\medskip

Moreover, the volatility seems to be specific in the sense that
\begin{itemize}
\item using other variables does not helps to anticipate volatility,
\item the quality of the anticipation of volatility dynamics does not vary a lot from one stock to another.
\end{itemize}
These two remarks could suggest that, once the seasonality is removed, volatility has a macroscopic (systematic) meaning, quite disconnected from the dynamics of other liquidity variables.
\medskip

In terms of the informational content of other variables:
\begin{itemize}
\item In most cases: using not only the variable itself, but the other ones to model its dynamics decreases the needed length of the past (i.e. the memory), without really improving the quality of the anticipation. 
\item The turnover dynamics helps to predict the volume of the book ones,
\item and all other variables dynamics are needed to improve the knowledge of bid-ask spread dynamics.
\item The UK seems to be specific in the sense the causality between turnover and volume of the book goes both ways: each of them significantly helps to reduce the number of lags needed to anticipate dynamics of the other.
\end{itemize}
\medskip

Our results clearly exhibit auto-correlations between variables driving liquidity dynamics at the five-minute time scale. Since most frameworks using for optimal trading and transaction cost analysis are positioned at this scale, it could be good to consider to introduce our endogenous dynamics into these frameworks. This opens a new direction of research.